\documentclass[a4paper,11pt]{article}

\usepackage{amsmath,amssymb,bbm,mathtools}
\usepackage{ascmac}
\usepackage{cancel}
\usepackage{xcolor} 
\usepackage{multirow}
\usepackage{marvosym,wasysym}
\usepackage[bookmarks=true,bookmarksnumbered=true,bookmarkstype=toc]{hyperref} 
\hypersetup{
pdfauthor={Tetsuji Kimura},
colorlinks={true},
linkcolor={blue},
urlcolor={blue},
filecolor={blue},
citecolor={blue}
}
\definecolor{refkey}{rgb}{0.40, 0.55, 0.55}
\definecolor{labelkey}{rgb}{0.40, 0.55, 0.55}
\usepackage{ulem}

\makeatletter

 \@addtoreset{equation}{section}
\makeatother

\newcounter{Enumerate}

\DeclareFontFamily{U}{rsf}{}
\DeclareFontShape{U}{rsf}{m}{n}{
  <5> <6> rsfs5 <7> <8> <9> rsfs7 <10-> rsfs10}{}
\DeclareMathAlphabet\Scr{U}{rsf}{m}{n}

\usepackage[mathscr]{eucal}

\newcommand{\del}{\partial}

\newcommand{\half}{\frac{1}{2}}

\newcommand{\ls}{\ \ \ \ \ }
\newcommand{\wt}{\widetilde}
\newcommand{\wh}{\widehat}

\newcommand{\ol}{\overline}

\newcommand{\bsubeq}{\begin{subequations}}
\newcommand{\esubeq}{\end{subequations}}

\newcommand{\noi}{\noindent}
\newcommand{\mr}{\mathring}

\newcommand{\eps}{\epsilon}
\newcommand{\nn}{\nonumber}

\newcommand{\I}{{\rm i}}
\newcommand{\N}{{\cal N}}
\renewcommand{\d}{{\rm d}}
\newcommand{\e}{{\rm e}}

\newcommand{\slb}{\scalebox}

\def\+{{+\!\!\!+}} 

\begin{document}
\allowdisplaybreaks{

\thispagestyle{empty}


\begin{flushright}
TIT/HEP-648 
\end{flushright}

\vspace{35mm}

\noi
\slb{2.1}{Semi-doubled Sigma Models for Five-branes}

%

\vspace{15mm}

\noi
{\renewcommand{\arraystretch}{1.6}
\begin{tabular}{cl}
\multicolumn{2}{l}{\slb{1.2}{Tetsuji {\sc Kimura}} \vphantom{$\Bigg|$}}
\\
& {\renewcommand{\arraystretch}{1.0}
\begin{tabular}{l}
{\sl Research and Education Center for Natural Sciences, Keio University}
\\
{\sl Hiyoshi 4-1-1, Yokohama, Kanagawa 223-8521, JAPAN} 
\end{tabular}
}
\\
& \ls and
\\
& {\renewcommand{\arraystretch}{1.0}
\begin{tabular}{l}
{\sl
Department of Physics,
Tokyo Institute of Technology} \vphantom{$\Big|$}
\\
{\sl Tokyo 152-8551, JAPAN}
\end{tabular}
}
\\
& \ \ \ \slb{0.9}{\tt tetsuji.kimura \_at\_ keio.jp}
\end{tabular}
}

\vspace{20mm}


\slb{1.1}{\sc Abstract}
\begin{center}
\slb{.95}{
\begin{minipage}{.95\textwidth}
\parindent=6mm%
We study two-dimensional $\N=(2,2)$ gauge theory and its dualized system in terms of complex (linear) superfields and their alternatives.
Although this technique itself is not new,
we can obtain a new model, the so-called ``semi-doubled'' GLSM.
Similar to doubled sigma model,
this involves both the original and dual degrees of freedom simultaneously,
whilst the latter only contribute to the system via topological interactions.
Applying this to the $\N=(4,4)$ GLSM for H-monopoles, i.e., smeared NS5-branes,
we obtain its T-dualized systems in quite an easy way.
As a bonus, we also obtain the semi-doubled GLSM for an exotic $5^3_2$-brane whose background is locally nongeometric.
In the low energy limit, we construct the semi-doubled NLSM which also generates the conventional string worldsheet sigma models.
In the case of the NLSM for $5^3_2$-brane, however, 
we find that the Dirac monopole equation does not make sense any more because the physical information is absorbed into the divergent part via the smearing procedure.
This is nothing but the signal which indicates that the nongeometric feature emerges in the considering model.

\end{minipage}
}
\end{center}


\newpage
\section{Introduction}
\label{sect:intro}

In string theory there are a lot of extended objects: a fundamental string,
an NS5-brane and D-branes in ten dimensions.
Performing string dualities in lower dimensional spacetime, 
we encounter different kind of objects, called exotic branes 
\cite{Elitzur:1997zn, Blau:1997du, Obers:1998fb, Eyras:1999at}.
The exotic brane is of codimension less than three, and its tension is often stronger than those of ordinary branes.
These days the exotic branes have been exhaustively investigated in the framework of 
supergravity theories 
\cite{Hellerman:2002ax, deBoer:2010ud, deBoer:2012ma, Kimura:2014wga, Okada:2014wma, Kimura:2014bea, Park:2015gka, Lust:2015yia},
string worldsheet theories
\cite{Kawai:2007qd, Kikuchi:2012za, Kimura:2013fda, Kimura:2013zva, Kimura:2013khz, Kimura:2014aja, Kimura:2015yla},
worldvolume theories
\cite{Bergshoeff:2006jj, Bergshoeff:2010xc, Bergshoeff:2011zk, Bergshoeff:2012pm, Chatzistavrakidis:2013jqa, Kimura:2014upa},
extended geometries such as 
doubled sigma model and double field theory
\cite{Albertsson:2008gq, Albertsson:2011ux, Hatsuda:2012uk, Hassler:2013wsa, Berkeley:2014nza, Berman:2014jsa, Ho:2014una, Ma:2014kia, Ma:2014vqm, Berman:2014hna, Ma:2015yma},
$\beta$-supergravity and its extension
\cite{Andriot:2014uda, Blair:2014zba, Sakatani:2014hba},
and other huge number of related works\footnote{Of course this classification is not rigorous because they are deeply related to each other.}.

The exotic $5^2_2$-brane originates from the NS5-brane via T-duality along two directions of the four-dimensional transverse space.
Then the four transverse directions of the $5^2_2$-brane is written as a two-torus fibration of a two-dimensional plane.
The exotic structure is that the transition function of this geometry is governed by not only the coordinate transformation group but also the T-duality group.
This implies that, going around the five-brane, the size of the torus is T-dualized, and the background geometry becomes multi-valued.
In order to understand this feature from the viewpoint of the string worldsheet,
the author has investigated mainly the exotic $5^2_2$-brane in the framework of two-dimensional supersymmetric gauge theory in \cite{Kimura:2013fda}.
This model is referred to as the gauged linear sigma model (or GLSM, for short) \cite{Witten:1993yc}.
The GLSM is the UV completion of the nonlinear sigma model (NLSM), which provides us the string worldsheet theory.

The previous work \cite{Kimura:2013fda} is motivated by the developed works \cite{Tong:2002rq, Harvey:2005ab, Okuyama:2005gx} in the language of the $\N=(4,4)$ GLSM,
where the target space configuration of the low energy effective NLSM is NS5-branes or KK-monopoles.
Applying further T-duality to this,
the GLSM description of the exotic $5^2_2$-brane was successfully obtained \cite{Kimura:2013fda}.
However, the procedure to derive such a model is technically complicated.
This is because the duality transformation of superfields in the model is performed only in terms of $\N=(2,2)$ irreducible superfields.
Then, the construction of the first order Lagrangian which derives the original GLSM and its dual system requires introducing many auxiliary superfields, most of which are just integrated out in the process of the duality transformation.

In this paper, we continue to construct a more useful and powerful model than the previous one.
It is known that the duality transformations without isometry can be performed in terms of complex linear superfields \cite{Gates:1984nk} (for instance, see the review \cite{Grisaru:1997ep}).
We expect that this will leads us to some faithful features of (non)geometric structures.
Actually, applying this technique to the $\N=(4,4)$ GLSM \cite{Tong:2002rq},
we will be able to construct the model for the exotic $5^2_2$-brane and for the exotic $5^3_2$-brane.
The latter is regarded as genuinely exotic because its background geometry is even locally nongeometric.
We would like to extract such a nongeometric feature in the framework of string worldsheet sigma model and its UV completion.

\newpage

The structure of this paper is as follows.
In section \ref{sect:duality}, we briefly discuss the duality transformation without isometry.
First, we find a dualized Lagrangian with the duality relation between a (twisted) chiral superfield and a complex (twisted) linear superfield.
We notice that the former is irreducible, while the latter reducible.
Next, we replace the complex (twisted) linear superfield to the sum of irreducible superfields.
This is an important preliminary to investigate the GLSM and its T-dualized systems, and their low energy effective theories.
In section \ref{sect:components},
we develop the duality transformation in terms of the component fields of the superfields.
There emerge various fields, some of them are redundant and integrated out.
We finally obtain a new first order Lagrangian, which we refer to as the ``semi-doubled'' Lagrangian. 
This provides us not only the original Lagrangian but also the dual Lagrangians.
In section \ref{sect:SD-GLSM},
we study the ``semi-doubled'' GLSM for five-branes.
Applying the technique which we obtain in the previous section,
we obtain the conventional GLSM for H-monopoles (i.e., smeared NS5-branes), KK-monopoles, and an exotic $5^2_2$-brane in a straightforward way.
Further, we propose a semi-doubled GLSM for an exotic $5^3_2$-brane whose background is locally nongeometric.
In section \ref{sect:SD-NLSM},
We obtain the ``semi-doubled'' NLSM as the low energy effective theory of the semi-doubled GLSM.
Performing the duality transformations along certain directions,
we precisely realize the NLSM for the H-monopoles, the KK-monopoles, and the $5^2_2$-brane. 
However, we cannot obtain the consistent description of the $5^3_2$-brane because the Dirac monopole equation is broken down.
We conclude that this background is nongeometric.
Section \ref{sect:summary} is devoted to the summary. 
In appendix \ref{app:SF}, the conventions in this paper are exhibited.
In appendix \ref{app:DT}, we briefly discuss the duality transformation rules with(out) isometry in two-dimensional $\N=(2,2)$ theories.
This is based on the work by Grisaru, Massar, Sevrin and Troost \cite{Grisaru:1997ep}, Ro\v{c}ek and Verlinde \cite{Rocek:1991ps}, Hori and Vafa \cite{Hori:2000kt} and Tong \cite{Tong:2002rq}. 

\section{Duality transformations in superfield formalism}
\label{sect:duality}

In this section 
we discuss the duality transformation which interchanges a (twisted) chiral superfield for a complex (twisted) linear superfield in a concrete way.
We should notice that we can perform the duality transformation even without isometry, where isometry is broken by the existence of (twisted) F-term.
A generic discussion can be seen in appendix \ref{app:DT} which is based on \cite{Grisaru:1997ep}.

\subsection{Chiral superfields with F-term}
\label{sect:CSF-F}

Let us begin with a Lagrangian 
\begin{align}
\Scr{L}_{\Psi}
\ &= \ 
\int \d^4 \theta \, \frac{1}{g^2} |\Psi|^2
+ \Big\{
- \sqrt{2} \int \d^2 \theta \, \Psi \Phi
+ \text{(h.c.)}
\Big\}
\nn \\
\ &= \ 
\int \d^4 \theta \, \Big\{
\frac{1}{g^2} |\Psi|^2
- 2 \sqrt{2} \, \Psi C - 2 \sqrt{2} \, \ol{\Psi} \, \ol{C}
\Big\}
\, . \label{L-Psi}
\end{align}
Here $\Psi$ and $\Phi$ are $\N=(2,2)$ chiral superfields,
while $C$ is the prepotential\footnote{In this work we do not introduce the ``gauge-fixing'' condition discussed in \cite{Kimura:2015cza}.} of $\Phi$ defined by $\Phi = \ol{D}{}_+ \ol{D}{}_- C$.
We note that the conventions of superfields are exhibited in appendix \ref{app:SF}.
$g$ is a dimensionless sigma model coupling constant.
Due to the existence of the F-term $- \sqrt{2} \Psi \Phi$, this model has no isometry\footnote{Supersymmetric sigma models with F-term and their duality transformations only in terms of irreducible superfields were recently discussed in \cite{Kimura:2014aja}.}.
In order to consider the duality transformation, 
we introduce the first order Lagrangian of (\ref{L-Psi}) such as 
\begin{align}
\Scr{L}_{RLC}
\ &= \ 
\int \d^4 \theta \, \Big\{
\frac{1}{g^2} |R|^2
- 2 \sqrt{2} \, R C
- 2 \sqrt{2} \, \ol{R} \, \ol{C}
- R L 
- \ol{R} \, \ol{L}
\Big\}
\, , \label{L-RLC}
\end{align}
where $R$ is an unconstrained complex prepotential
and $L$ is a complex linear superfield whose definition is 
$0 = \ol{D}{}_+ \ol{D}{}_- L$.
This Lagrangian leads to two second order Lagrangians.
One is the original form (\ref{L-Psi}) and the other is the dualized form which we will show.

First, we evaluate the equation of motion for the complex linear superfield $L$.
This gives a constraint on the prepotential $R$ such as
$0 = \ol{D}{}_{\pm} R$.
Under this field equation $R$ is restricted to a chiral superfield $X$:
\begin{align}
R \ &= \ 
X
\, . \label{R-Psi}
\end{align}
Substituting this into (\ref{L-RLC})
and identifying $X$ with $\Psi$, we go back to the original Lagrangian (\ref{L-Psi}).
On the other hand, the equation of motion for the prepotential $R$ in (\ref{L-RLC}) is given as
\begin{align}
0 \ &= \ 
\frac{1}{g^2} \ol{R} - 2 \sqrt{2} \, C - L
\, . \label{R-LC}
\end{align}
Plugging this into (\ref{L-RLC}), we obtain
\begin{align}
\Scr{L}_{RLC}
\ &= \ 
- g^2 \int \d^4 \theta \, \Big| L + 2 \sqrt{2} \, C \Big|^2 
\ \equiv \ 
\Scr{L}_{LC}
\, . \label{L-LC}
\end{align}
This is the dualized Lagrangian from the original one (\ref{L-Psi}).
We find the duality relation between the original chiral superfield $\Psi$ and the dual complex linear superfield $L$
via (\ref{R-Psi}) and (\ref{R-LC}): 
\begin{align}
\frac{1}{g^2} \ol{\Psi} 
\ &= \ 
L + 2 \sqrt{2} \, C
\, . \label{Psi2LC}
\end{align}
We emphasize that the above duality transformation rule is quite simple and straightforward compared with those discussed in \cite{Kimura:2014aja}.

\subsection{Twisted chiral superfields with twisted F-term}
\label{sect:TCSF-TF}

Analogous to the duality transformation of the chiral superfield,
we consider the duality transformation of twisted chiral superfields with twisted F-term.
We start from a second order Lagrangian 
\begin{align}
\Scr{L}_{\Theta}
\ &= \ 
- \frac{1}{g^2} \int \d^4 \theta \, |\Theta|^2
+ \Big\{
- \sqrt{2} \, \int \d^2 \wt{\theta} \, \Theta \Sigma
+ \text{(h.c.)}
\Big\}
\nn \\
\ &= \
\int \d^4 \theta \, \Big\{
- \frac{1}{g^2} |\Theta|^2
- 2 (\Theta + \ol{\Theta}) V
\Big\}
\, , \label{L-Theta} 
\end{align}
where $\Theta$ and $\Sigma$ are twisted chiral superfields and $V$ is a real vector superfield related to $\Sigma$ such as $\Sigma = \frac{1}{\sqrt{2}} \ol{D}{}_- D_- V$.
Strictly speaking, there is a total derivative term in the second line in the right-hand side.
Here we just ignore it because this does not contribute to the dualization, whilst it will be explicitly described in due course.
This plays a significant role in quantum analysis \cite{Tong:2002rq, Harvey:2005ab, Okuyama:2005gx, Kimura:2013zva}.
Since $V$ is real, this Lagrangian has an isometry along the imaginary part of $\Theta$.
For later discussions, however, it is important to study the duality transformation with isometry in terms of a complex twisted linear superfield\footnote{In appendix \ref{app:DT-tc}, we describe the duality transformation from a twisted chiral superfield with isometry to a chiral superfield in a standard way.}.
We introduce the first order Lagrangian of (\ref{L-Theta}) in the following form:
\begin{align}
\Scr{L}_{\wt{R} \wt{L}V}
\ &= \ 
\int \d^4 \theta \, \Big\{
- \frac{1}{g^2} |\wt{R}|^2
- 2 (\wt{R} + \ol{\wt{R}}) V
- \wt{R} \wt{L}
- \ol{\wt{R}} \, \ol{\wt{L}}
\Big\}
\, . \label{L-RtLV}
\end{align}
Here $\wt{R}$ is an unconstrained complex prepotential and 
$\wt{L}$ is a complex twisted linear superfield defined by 
$0 = \ol{D}{}_+ D_- \wt{L}$.
Analyzing the equation of motion for $\wt{L}$ or $\wt{R}$, we obtain the original Lagrangian or its dual form, respectively.
First, we evaluate the equation of motion for $\wt{L}$.
This gives a constraint on the prepotential $\wt{R}$ such as
$0 = \ol{D}{}_+ \wt{R} = D_- \wt{R}$.
Hence $\wt{R}$ is reduced to a twisted chiral superfield $Y$:
\begin{align}
\wt{R} \ &= \ Y
\, . \label{R-Theta}
\end{align}
Plugging this into (\ref{L-RtLV}) with identification $Y = \Theta$,
we obtain the original form (\ref{L-Theta}).
If we evaluate the equation of motion for $\wt{R}$ in (\ref{L-RtLV}),
we find
\begin{align}
0 \ &= \ 
- \frac{1}{g^2} \ol{\wt{R}} - 2 V - \wt{L}
\, . \label{R-tLV}
\end{align}
Under this field equation, the Lagrangian is reduced to
\begin{align}
\Scr{L}_{\wt{R}\wt{L}V}
\ &= \ 
g^2 \int \d^4 \theta \, \Big| \wt{L} + 2 V \Big|^2
\ \equiv \ 
\Scr{L}_{\wt{L}V}
\, . \label{L-tLV}
\end{align}
This is the dual Lagrangian from the original one (\ref{L-Theta}).
Through the equations (\ref{R-Theta}) and (\ref{R-tLV}), 
we find the duality relation between the original twisted chiral superfield $\Theta$ and the dual complex twisted linear superfield $\wt{L}$ in the following way:
\begin{align}
- \frac{1}{g^2} \ol{\Theta}
\ &= \ 
\wt{L} + 2 V
\, . \label{Theta2tLV}
\end{align}

We remark that while the present transformation is the dualization without isometry,
this can be also applicable in the presence of isometry.
This is a kind of generalization of the duality transformation by Ro\v{c}ek and Verlinde \cite{Rocek:1991ps}, and Hori and Vafa \cite{Hori:2000kt}. 
Then, in later discussions, we will apply the duality transformed Lagrangians (\ref{L-LC}) and (\ref{L-tLV}) to the GLSM for H-monopoles (smeared NS5-branes) and its T-dualized systems \cite{Tong:2002rq, Harvey:2005ab, Okuyama:2005gx, Kimura:2013fda, Kimura:2015yla}.
Originally this has no isometry along three of four real scalar fields, which represent the transverse directions of the H-monopoles in ten-dimensional string theory.
However, smearing the directions without isometry discussed in \cite{Sen:1994wr, Elitzur:1997zn, Blau:1997du, Cherkis:2000cj, Cherkis:2001gm, deBoer:2010ud, Kikuchi:2012za}, we can geometrically perform T-duality consistent with the Buscher rule \cite{Buscher:1987sk}.
In order to argue the same physical situation,
it is better to replace the complex (twisted) linear superfields with certain alternatives given by irreducible superfields.

\subsection{Replacements}
\label{sect:replace}

In the previous subsection, we discussed the duality transformation which interchanges a (twisted) chiral superfields without isometry and a complex (twisted) linear superfields.
In later sections, we will apply this technique to the GLSM for H-monopoles and its T-dualized systems, and their low energy effective theories as string worldsheet sigma models.

We should keep in mind that the superfield formalism is not so appropriate to investigate geometrical structures of the systems.
Hence we expand all superfields in terms of their component fields.
Now we should notice that complex (twisted) linear superfields are reducible.
Then, even in terms of the component fields, systems given by the complex (twisted) linear superfields might not be well understood.
In this subsection, we replace a complex (twisted) linear superfield with the sum of the irreducible superfields such as chiral and twisted chiral superfields\footnote{A similar discussion was demonstrated in \cite{Kimura:2015cza}.}.


Recall that the definition of a complex linear superfield is $0 = \ol{D}{}_+ \ol{D}{}_- L$.
Now we replace $L$ with the sum of the irreducible superfields in such a way as
\begin{align}
L \ &= \ 
X + Y + \ol{Z}
\, , \label{L=XYZ}
\end{align}
where $X$ is a chiral superfield, while $Y$ and $Z$ are twisted chiral superfields. 
All of the irreducible superfields carry two off-shell complex scalar fields and two off-shell complex Weyl fermions.
We note that $L$ carries six off-shell complex bosons and six off-shell complex Weyl fermions (see appendix \ref{app:SF}).
We can also replace a complex twisted linear superfield $\wt{L}$ with the sum of the irreducible superfields such as 
\begin{align}
\wt{L} 
\ &= \ 
X' + Y' + \ol{W}{}'
\, . \label{tL=XYW}
\end{align}
Here $X'$ and $W'$ are chiral superfields, while $Y'$ is a twisted chiral superfield.
The right-hand side of (\ref{tL=XYW}) vanishes if the operator $\ol{D}{}_+ D_-$ acts on it.
This is consistent with the definition of $\wt{L}$.
Again, the number of the component fields in the right-hand side is equal to that of the left-hand side.


Now we apply the replacements (\ref{L=XYZ}) and (\ref{tL=XYW}) to the dualized Lagrangians and the duality relations in the previous subsection:
\bsubeq \label{SF-relations}
\begin{align}
\Scr{L}_{LC}
\ &= \ 
- g^2 \int \d^4 \theta \, \Big| X + Y + \ol{Z} + 2 \sqrt{2} \, C \Big|^2
\, , \label{L-XYZC} \\
\Scr{L}_{\wt{L}V}
\ &= \ 
g^2 \int \d^4 \theta \, \Big| X' + Y' + \ol{W}{}' + 2 V \Big|^2
\, , \label{L-XYWV} \\
\frac{1}{g^2} \ol{\Psi}
\ &= \ 
X + Y + \ol{Z} + 2 \sqrt{2} \, C
\, , \label{Psi2XYZC} \\
- \frac{1}{g^2} \ol{\Theta}
\ &= \ 
X' + Y' + \ol{W}{}' + 2 V
\, . \label{Theta2XYWV}
\end{align}
\esubeq
Due to the replacements,
one might think that the duality relations (\ref{Psi2XYZC}) and (\ref{Theta2XYWV}) are inconsistent.
This is because the right-hand sides of (\ref{Psi2XYZC}) and (\ref{Theta2XYWV}) carry the degrees of freedom three times as many as those of the left-hand sides.
This is true.
However, we can remove redundant degrees of freedom in an appropriate way.
Furthermore, this ``unbalanced'' situation will lead us to a simple description of nongeometric background feature in the dualized system. 

Here we roughly mention the reduction of the redundant degrees of freedom, though in later discussions we will demonstrate it concretely: 
Focus on the dynamical scalar fields, i.e., six real scalars.
Two real bosons are replaced by the other bosonic degrees of freedom when we expand the duality relation in terms of the component fields.
Further two real bosons are decoupled from the system because they do not contribute to the system at all.
The remaining two real bosons are genuinely the dual degrees of freedom.

\section{Duality transformations by component fields}
\label{sect:components}

In this section we carefully investigate the dualized Lagrangians and the duality transformation rules (\ref{SF-relations}) in terms of the component fields of the various superfields.
As mentioned before, there exist many fields as the original fields, the dual fields and the redundant fields.
Compared with the duality transformation with isometry exhibited in appendix \ref{app:DT-tc}, we will determine which fields are redundant and integrated out.

\subsection{Expansions}
\label{sect:expansions}

We prepare the component fields of the superfields in (\ref{L-Theta}) and (\ref{SF-relations}).
Following the generic forms (\ref{expand}), we introduce the following expansions:
\bsubeq \label{ex-SF}
\begin{align}
\Psi \ &= \ 
\frac{1}{\sqrt{2}} (r^1 + \I r^2)
+ \I \sqrt{2} \, \theta^+ \chi_+
+ \I \sqrt{2} \, \theta^- \chi_-
+ 2 \I \, \theta^+ \theta^- G
+ \ldots 
\, , \label{Psi} \\
\Theta \ &= \ 
\frac{1}{\sqrt{2}} (r^3 + \I \vartheta)
+ \I \sqrt{2} \, \theta^+ \ol{\wt{\chi}}{}_+
- \I \sqrt{2} \, \ol{\theta}{}^- \wt{\chi}_-
+ 2 \I \, \theta^+ \ol{\theta}{}^- \wt{G}
+ \ldots 
\, , \label{Theta} \\
X \ &= \
\frac{1}{\sqrt{2}} (\phi_{X,1} + \I \phi_{X,2})
+ \I \sqrt{2} \, \theta^+ \psi_{X+} 
+ \I \sqrt{2} \, \theta^- \psi_{X-}
+ 2 \I \, \theta^+ \theta^- F_X
+ \ldots
\, , \label{X} \\
Y \ &= \
\frac{1}{\sqrt{2}} (\sigma_{Y,1} + \I \sigma_{Y,2})
+ \I \sqrt{2} \, \theta^+ \ol{\chi}{}_{Y+} 
- \I \sqrt{2} \, \ol{\theta}{}^- \chi_{Y-} 
+ 2 \I \, \theta^+ \ol{\theta}{}^- G_Y
+ \ldots
\, , \label{Y} \\
\ol{Z} \ &= \ 
\frac{1}{\sqrt{2}} (\sigma_{Z,1} - \I \sigma_{Z,2})
+ \I \sqrt{2} \, \ol{\theta}{}^+ \wt{\chi}_{Z+}
- \I \sqrt{2} \, \theta^- \ol{\wt{\chi}}{}_{Z-}
+ 2 \I \ol{\theta}{}^+ \theta^- \ol{\wt{G}}{}_Z
+ \ldots
\, , \label{Z} \\
X' \ &= \ 
\frac{1}{\sqrt{2}} (\phi'_{X,1} + \I \phi'_{X,2})
+ \I \sqrt{2} \, \theta^+ \psi'_{X+} 
+ \I \sqrt{2} \, \theta^- \psi'_{X-}
+ 2 \I \, \theta^+ \theta^- F'_X
+ \ldots
\, , \label{X'} \\
Y' \ &= \ 
\frac{1}{\sqrt{2}} (\sigma'_{Y,1} + \I \sigma'_{Y,2})
+ \I \sqrt{2} \, \theta^+ \ol{\chi}{}'_{Y+} 
- \I \sqrt{2} \, \ol{\theta}{}^- \chi'_{Y-} 
+ 2 \I \, \theta^+ \ol{\theta}{}^- G'_Y
+ \ldots
\, , \label{Y'} \\
\ol{W}{}' \ &= \ 
\frac{1}{\sqrt{2}} (\phi'_{W,1} - \I \phi'_{W,2})
+ \I \sqrt{2} \, \ol{\theta}{}^+ \ol{\psi}{}'_{W+}
+ \I \sqrt{2} \, \ol{\theta}{}^- \ol{\psi}{}'_{W-} 
+ 2 \I \, \ol{\theta}{}^+ \ol{\theta}{}^- \ol{F}{}'_W 
+ \ldots
\, . \label{W'}
\end{align}
\esubeq
The expansions of $V$ and $C$ are expressed in (\ref{V}) and (\ref{C}), respectively.
Each superfield starts from a pair of two real scalar fields.
The second and third terms contain the fermionic fields as complex Weyl spinors whose subscripts $\pm$ represent their chirality.
The fourth term in each superfield represents the auxiliary field as a complex scalar.
The terms ``$\ldots$'' involve derivative terms.

Let us substitute (\ref{ex-SF}) into the data in the previous section.
From now on, we just ignore the fermionic degrees of freedom because we can restore them via the supersymmetry transformations.
First, we evaluate the original second order Lagrangians (\ref{L-Psi}) and (\ref{L-Theta}):
\bsubeq \label{L-original-comp}
\begin{align}
\Scr{L}_{\Psi}
\ &= \ 
- \frac{1}{2 g^2} (\del_m r^1)^2
- \frac{1}{2 g^2} (\del_m r^2)^2
+ \frac{1}{g^2} |G|^2
- \sqrt{2} \, (G M_{c} + \ol{G} \ol{M}{}_{c})
\nn \\
\ & \ \ \ \ 
- r^1 (D_{c} + \ol{D}{}_{c})
+ \frac{1}{2} \del_+ \del_- r^1 (\phi_{c} + \ol{\phi}{}_{c})
+ \frac{\I}{2} \del_+ r^1 (A_{c=} - \ol{A}{}_{c=})
+ \frac{\I}{2} \del_- r^1 (B_{c\+} - \ol{B}{}_{c\+})
\nn \\
\ & \ \ \ \ 
- \I r^2 (D_{c} - \ol{D}{}_{c})
+ \frac{\I}{2} \del_+ \del_- r^2 (\phi_{c} - \ol{\phi}{}_{c})
- \frac{1}{2} \del_+ r^2 (A_{c=} + \ol{A}{}_{c=})
- \frac{1}{2} \del_- r^2 (B_{c\+} + \ol{B}{}_{c\+})
\, , \label{L-PsiC-comp} \\
\Scr{L}_{\Theta}
\ &= \ 
- \frac{1}{2 g^2} (\del_m r^3)^2
- \frac{1}{2 g^2} (\del_m \vartheta)^2
+ \frac{1}{g^2} |\wt{G}|^2
+ \sqrt{2} \Big\{
  r^3 D_{V} + \vartheta \, F_{01}
- \I \, \sigma \wt{G} + \I \, \ol{\sigma} \ol{\wt{G}}{}
\Big\}
\, , \label{L-Theta-comp}
\end{align}
\esubeq
where the gauge field strength is defined as $F_{01} = \eps^{mn} \del_m A_n$ by virtue of the invariant tensor whose normalization is $\eps^{01} = - \eps^{10} = + 1$.
Next, we expand the dual Lagrangians (\ref{L-LC}) and (\ref{L-tLV}):
\bsubeq \label{L-dual-comp}
\begin{align}
\Scr{L}_{LC}
\ &= \ 
- \frac{g^2}{2}
\Big\{ (\del_m \sigma_{Y,1})^2 + (\del_m \sigma_{Z,1})^2 \Big\}
+ \frac{g^2}{2} (\del_m \phi_{X,1})^2
\nn \\
\ & \ \ \ \ 
- \frac{g^2}{2}
\Big\{ (\del_m \sigma_{Y,2})^2 + (\del_m \sigma_{Z,2})^2 \Big\}
+ \frac{g^2}{2} (\del_m \phi_{X,2})^2
\nn \\
\ & \ \ \ \ 
- 2 g^2 |M_{c}|^2
- g^2 \Big| F_X + \sqrt{2} \, F_c \Big|^2
+ g^2 \Big| \I \wt{G}_Y + \sqrt{2} \, G_c \Big|^2
+ g^2 \Big| \I \wt{G}_Z + \sqrt{2} \, \ol{N}{}_c \Big|^2
\nn \\
\ & \ \ \ \ 
- g^2 \Big\{ \phi_{X,1} + (\sigma_{Y,1} + \sigma_{Z,1}) + 2 \sqrt{2} \, \phi_{c,1} \Big\} (D_c + \ol{D}{}_c)
\nn \\
\ & \ \ \ \ 
+ \frac{g^2}{2} \, \del_+ \del_- \big( \phi_{X,1} - (\sigma_{Y,1} + \sigma_{Z,1}) \big) (\phi_{c} + \ol{\phi}{}_{c})
+ g^2 (A_{c=} - \ol{A}{}_{c=}) (B_{c\+} - \ol{B}{}_{c\+})
\nn \\
\ & \ \ \ \ 
- \frac{\I g^2}{2} \, \del_+ \big( \phi_{X,1} + (\sigma_{Y,1} - \sigma_{Z,1}) \big) (A_{c=} - \ol{A}{}_{c=})
%
- \frac{\I g^2}{2} \, \del_- \big( \phi_{X,1} - (\sigma_{Y,1} - \sigma_{Z,1}) \big) (B_{c\+} - \ol{B}{}_{c\+})
\nn \\
\ & \ \ \ \ 
+ \I g^2 \Big\{ \phi_{X,2} + (\sigma_{Y,2} - \sigma_{Z,2}) + 2 \sqrt{2} \, \phi_{c,2} \Big\} (D_c - \ol{D}{}_c)
\nn \\
\ & \ \ \ \ 
- \frac{\I g^2}{2} \, \del_+ \del_- \big( \phi_{X,2} - (\sigma_{Y,2} - \sigma_{Z,2}) \big) (\phi_{c} - \ol{\phi}{}_{c})
- g^2 (A_{c=} + \ol{A}{}_{c=}) (B_{c\+} + \ol{B}{}_{c\+})
\nn \\
\ & \ \ \ \ 
- \frac{g^2}{2} \,
\del_+ \big( \phi_{X,2} + (\sigma_{Y,2} + \sigma_{Z,2}) \big) (A_{c=} + \ol{A}{}_{c=})
%
- \frac{g^2}{2} \, \del_- \big( \phi_{X,2} - (\sigma_{Y,2} + \sigma_{Z,2}) \big) (B_{c\+} + \ol{B}{}_{c\+})
\, , \label{L-LC-comp} \\
\Scr{L}_{\wt{L}V}
\ &= \ 
- \frac{g^2}{2}
\Big\{
  (\del_m \phi'_{X,1})^2
+ (\del_m \phi'_{W,1})^2
\Big\}
+ \frac{g^2}{2} (\del_m \sigma'_{Y,1})^2
\nn \\
\ & \ \ \ \ 
- \frac{g^2}{2}
\Big\{
  (D_m \phi'_{X,2})^2
+ (D_m \phi'_{W,2})^2
\Big\}
+ \frac{g^2}{2} (\del_m \sigma'_{Y,2})^2
+ \sqrt{2} \, g^2 \eps^{mn} (\del_m \sigma'_{Y,2}) A_n
\nn \\
\ & \ \ \ \ 
- 4 g^2 |\sigma|^2 
+ \I \sqrt{2} \, g^2 \Big( \sigma \wt{G}{}'_Y - \ol{\sigma} \ol{\wt{G}}{}'_Y \Big)
+ g^2 \Big( |F'_X|^2 + |F'_W|^2 - |\wt{G}{}'_Y|^2 \Big)
\nn \\
\ & \ \ \ \ 
- \sqrt{2} \, g^2 D_{V} \big( \phi'_{X,1} + \phi'_{W,1} + \sigma'_{Y,1} \big)
\, . \label{L-tLV-comp}
\end{align}
\esubeq
Here we introduced the gauge covariant derivatives
\begin{alignat}{2}
D_m \phi'_{X,2}
\ &= \ 
\del_m \phi'_{X,2} - \sqrt{2} A_m
\, , &\ls
D_m \phi'_{W,2}
\ &= \ 
\del_m \phi'_{W,2} - \sqrt{2} A_m
\, . \label{gauge-cov}
\end{alignat}
Finally, we describe the duality relations (\ref{Psi2XYZC}) and (\ref{Theta2XYWV}) in terms of the component fields.
The duality relation (\ref{Psi2XYZC}) provides the following relations:
\bsubeq \label{Psi2XYZC-comp}
\begin{align}
\frac{1}{g^2} r^1
\ &= \ 
+ \big( \phi_{X,1} + (\sigma_{Y,1} + \sigma_{Z,1}) \big) 
+ 2 \sqrt{2} \, \phi_{c,1}
\, , \label{r1-L} \\
\frac{1}{g^2} r^2
\ &= \ 
- \big( \phi_{X,2} + (\sigma_{Y,2} - \sigma_{Z,2}) \big) 
- 2 \sqrt{2} \, \phi_{c,2}
\, , \label{r2-L} \\
\frac{1}{g^2} \del_m r^1 
\ &= \ 
- \del_m \phi_{X,1}
+ \eps_{mn} \del^n (\sigma_{Y,1} - \sigma_{Z,1})
- 2 \I (W_{c,m} - \ol{W}{}_{c,m})
\, , \label{dr1-L} \\
\frac{1}{g^2} \del_m r^2
\ &= \ 
+ \del_m \phi_{X,2}
- \eps_{mn} \del^n (\sigma_{Y,2} + \sigma_{Z,2})
+ 2 (W_{c,m} + \ol{W}{}_{c,m})
\, , \label{dr2-L} \\
\frac{1}{g^2} \ol{G}
\ &= \ 
\sqrt{2} \, M_{c}
\, , \\
0 \ &= \ 
- \I \ol{\wt{G}}{}_Z + \sqrt{2} \, N_{c}
\, , \\
0 \ &= \ 
+ \I \wt{G}_Y + \sqrt{2} \, G_{c}
\, , \\
0 \ &= \ 
F_X + \sqrt{2} \, F_{c}
\, .
\end{align}
\esubeq
Here we introduced a complex vector field $W_{c,m}$ with $A_{c=} = W_{c,0} - W_{c,1}$ and $B_{c\+} = W_{c,0} + W_{c,1}$, because $A_{c=}$ and $B_{c\+}$ are complex vectorial fields.
The relation (\ref{r1-L}) denotes that $\sigma_{Y,1} + \sigma_{Z,1}$ seems to be the original field $r^1$, while (\ref{dr1-L}) implies that $\sigma_{Y,1} - \sigma_{Z,1}$ behaves as the dual field of $r^1$.
The relations (\ref{r2-L}) and (\ref{dr2-L}) also indicate that $\sigma_{Y,2} - \sigma_{Z,2}$ is the same as the original field $r^2$, while $\sigma_{Y,2} + \sigma_{Z,2}$ as the dual field.
As we will see later, 
the fields $(\phi_{X,1}, \phi_{X,2})$ play a distinctive role in the dual system. 
In the same way, we can read off the duality relations among the component fields from (\ref{Theta2XYWV}):
\bsubeq \label{Theta2XYWV-comp}
\begin{align}
\frac{1}{g^2} r^3 
\ &= \ 
- \big( (\phi'_{X,1} + \phi'_{W,1}) + \sigma'_{Y,1} \big)
\, , \label{r3-tL} \\
\frac{1}{g^2} \vartheta
\ &= \ 
+ \big( (\phi'_{X,2} - \phi'_{W,2}) + \sigma'_{Y,2} \big)
\, , \label{r4-tL} \\
\frac{1}{g^2} \del_m r^3 
\ &= \ 
- \eps_{mn} \del^n \big( \phi'_{X,1} - \phi'_{W,1} \big)
+ \del_m \sigma'_{Y,1}
\, , \label{dr3-tL} \\
\frac{1}{g^2} \del_m \vartheta
\ &= \ 
+ \eps_{mn} D^n \big( \phi'_{X,2} + \phi'_{W,2} \big)
- \del_m \sigma'_{Y,2}
\, , \label{dr4-tL} \\
- \frac{\I}{g^2} \ol{\wt{G}}
\ &= \ 
\sqrt{2} \, \sigma
\, , \\
0 \ &= \ 
F'_X
\, , \\
0 \ &= \ 
\ol{F}{}'_W 
\, , \\
0 \ &= \ 
\I \wt{G}'_Y - \sqrt{2} \, \ol{\sigma}
\, .
\end{align}
\esubeq
Here we omitted the duality relations among the fermionic fields.
The relations (\ref{r3-tL}), (\ref{r4-tL}), (\ref{dr3-tL}) and (\ref{dr4-tL}) give us the following interpretations: 
$\phi'_{X,1} + \phi'_{W,1}$ and $\phi'_{X,2} - \phi'_{W,2}$ correspond to the original fields $r^3$ and $\vartheta$, respectively, 
whilst $\phi'_{X,1} - \phi'_{W,1}$ and $\phi'_{X,2} + \phi'_{W,2}$ are the dual fields of $r^3$ and $\vartheta$.
On the other hand, as discussed later,
$\sigma'_{Y,1}$ and $\sigma'_{Y,2}$ are not canonical fields.

\subsection{Eliminating redundant fields}
\label{sect:int-out}

We now investigate the Lagrangians (\ref{L-dual-comp}).
They contain many redundant fields which should be eliminated by virtue of the duality relations (\ref{Psi2XYZC-comp}) and (\ref{Theta2XYWV-comp}).
The strategy is as follows:
In the beginning, we find that terms of auxiliary fields are simplified by the duality relations.
Second, we focus on a field whose kinetic term is not canonical.
We eliminate it via the duality relations in order that the reduced Lagrangian can generate both the original and the dual ones when we integrate out certain dynamical fields.
Finally, we integrate out fields which do not contribute to the system at all.

\subsubsection*{Lagrangian $\Scr{L}_{\wt{L}V}$} 

Let us first consider the Lagrangian $\Scr{L}_{\wt{L}V}$ (\ref{L-tLV-comp}) 
dualized from the original one $\Scr{L}_{\Theta}$ (\ref{L-Theta-comp}). 
Since $\Scr{L}_{\Theta}$ has an isometry along $\vartheta$,
we can dualize it in a standard way as $\Scr{L}_{\Gamma V}$ (\ref{L-GV-comp-app}) in appendix \ref{app:DT-tc}.
Thus we should keep in mind that (\ref{L-tLV-comp}) derives the same form as (\ref{L-GV-comp-app}). 

In order to make (\ref{L-tLV-comp}) simple,
we introduce the following expressions:
\begin{alignat}{2}
\phi'_{1\pm} 
\ &\equiv \ 
\phi'_{X,1} \pm \phi'_{W,1}
\, , &\ls
\phi'_{2\pm}
\ &\equiv \ 
\phi'_{X,2} \pm \phi'_{W,2}
\, . \label{re-sigma12}
\end{alignat}
Then the covariant derivatives of $(\phi'_{X,2}, \phi'_{W,2})$ are combined into $D_m \phi'_{2+} = \del_m \phi'_{2+} - 2 \sqrt{2} A_m$.
Substituting (\ref{re-sigma12}) and the relations among the auxiliary fields $(F'_X, \wt{G}'_Y, F'_W)$ from (\ref{Theta2XYWV-comp}) into the Lagrangian (\ref{L-tLV-comp}), we find
\begin{align}
\Scr{L}_{\wt{L}V}
\ &= \ 
- \frac{g^2}{4}
\Big\{
  (\del_m \phi'_{1+})^2
+ (\del_m \phi'_{1-})^2
\Big\}
+ \frac{g^2}{2} (\del_m \sigma'_{Y,1})^2
\nn \\
\ & \ \ \ \ 
- \frac{g^2}{4}
\Big\{
  (D_m \phi'_{2+})^2
+ (D_m \phi'_{2-})^2
\Big\}
+ \frac{g^2}{2} (\del_m \sigma'_{Y,2})^2
+ \sqrt{2} \, g^2 \eps^{mn} (\del_m \sigma'_{Y,2}) A_n
\nn \\
\ & \ \ \ \ 
- 2 g^2 |\sigma|^2 
- \sqrt{2} \, g^2 D_{V} \big( \phi'_{1+} + \sigma'_{Y,1} \big)
\, . \label{L-tLV-comp2}
\end{align}
We immediately find that the kinetic terms of $\sigma'_{Y,1}$ and $\sigma'_{Y,2}$ are not canonical.
Then we eliminate them by virtue of the duality relations (\ref{Theta2XYWV-comp}).
We symbolically express the derivatives of (\ref{r3-tL}) and of (\ref{r4-tL}),
the relations (\ref{dr3-tL}) and (\ref{dr4-tL}) themselves, and the kinetic terms as follows:
\bsubeq \label{sigmaY12-abc}
\begin{align}
\del_m \wt{\sigma}'_{Y,1}
\ &\equiv \ 
- \frac{1}{g^2} \del_m r^3
- \del_m \phi'_{1+}
\, , \\
\del_m \wh{\sigma}'_{Y,1}
\ &\equiv \ 
+ \frac{1}{g^2} \del_m r^3
+ \eps_{mn} \del^n \phi'_{1-}
\, , \\
\del_m \wt{\sigma}'_{Y,2}
\ &\equiv \ 
+ \frac{1}{g^2} \del_m \vartheta
- \del_m \phi'_{2-}
\, , \\
\del_m \wh{\sigma}'_{Y,2}
\ &\equiv \ 
- \frac{1}{g^2} \del_m \vartheta
+ \eps_{mn} D^n \phi'_{2+}
\, , \\
+ \frac{g^2}{2} (\del_m \sigma'_{Y,1})^2
\ &\equiv \ 
\frac{g^2}{4} \Big\{
(\del_m \wt{\sigma}'_{Y,1})^2
- (\del_m \wh{\sigma}'_{Y,1})^2
+ 2 (\del_m \wt{\sigma}'_{Y,1}) (\del^m \wh{\sigma}'_{Y,1})
\Big\}
\, , \label{kin-sigmaY1} \\
+ \frac{g^2}{2} (\del_m \sigma'_{Y,2})^2
\ &\equiv \ 
\frac{g^2}{4} \Big\{
(\del_m \wt{\sigma}'_{Y,2})^2
- (\del_m \wh{\sigma}'_{Y,2})^2
+ 2 (\del_m \wt{\sigma}'_{Y,2}) (\del^m \wh{\sigma}'_{Y,2})
\Big\}
\, . \label{kin-sigmaY2} 
\end{align}
\esubeq
While we substitute only (\ref{r3-tL}) and (\ref{r4-tL}) into the interaction terms.
We are now ready to remove the ``tachyonic'' fields $\sigma'_{Y,1}$ and $\sigma'_{Y,2}$. 
Combining the above result into the Lagrangian (\ref{L-tLV-comp2}),
we obtain
\begin{align}
\Scr{L}_{\wt{L}V}
\ &= \ 
- \frac{1}{2 g^2} (\del_m r^3)^2
- \eps^{mn} (\del_m r^3) (\del_n \phi'_{1-})
- \frac{g^2}{2} \eps^{mn} (\del_m \phi'_{1+}) (\del_n \phi'_{1-})
\nn \\
\ & \ \ \ \ 
- \frac{1}{2 g^2} (\del_m \vartheta)^2
+ \eps^{mn} (\del_m \vartheta) (D_n \phi'_{2+})
- \frac{g^2}{2} \eps^{mn} (\del_m \phi'_{2-}) (D_n \phi'_{2+})
\nn \\
\ & \ \ \ \
+ \sqrt{2} \, \eps^{mn} A_n (\del_m \vartheta)
- \sqrt{2} \, g^2 \eps^{mn} (\del_m \phi'_{2-}) A_n 
- 2 g^2 |\sigma|^2 + \sqrt{2} \, r^3 D_{V}
\, . \label{L-tLV-comp3}
\end{align}
This is not the end of story.
We see that $\phi'_{1+}$ and $\phi'_{2-}$ do not couple to the original scalar fields $r^3$ and $\vartheta$.
Then it is possible to integrate them out.
Since their equations of motion are trivially satisfied by virtue of the invariant tensor $\eps^{mn}$, we can simply remove them away.
Then the final form is given as follows:
\begin{align}
\Scr{L}_{\wt{L}V}
\ &= \ 
- \frac{1}{2 g^2} (\del_m r^3)^2
- \frac{1}{2 g^2} (\del_m \vartheta)^2
- \eps^{mn} (\del_m r^3) (\del_n \phi'_{1-})
+ \eps^{mn} (\del_m \vartheta) (D_n \phi'_{2+})
\nn \\
\ & \ \ \ \ 
+ \sqrt{2} \, \eps^{mn} (\del_m \vartheta) A_n 
- 2 g^2 |\sigma|^2 + \sqrt{2} \, r^3 D_{V}
\, . \label{L-tLV-comp4}
\end{align}
This Lagrangian involves both the original fields $(r^3, \vartheta)$ and their dual fields $(\phi'_{1-}, \phi'_{2+})$, 
while the dual ones do not have kinetic terms explicitly.
However, we can correctly derive the dual Lagrangian if we integrate out the original fields.
Focus on the $(\vartheta, \phi'_{2+})$ sector.
Evaluating the equation of motion for $\vartheta$, we obtain 
\begin{align}
\del_m \vartheta
\ &= \ 
g^2 \eps_{mn} D^n \phi'_{2+}
+ \sqrt{2} \, g^2 \eps_{mn} A^n
\, .
\end{align}
Plugging this into the $(\vartheta, \phi'_{2+})$ sector in (\ref{L-tLV-comp4}), 
we obtain the dual form in $\Scr{L}_{\Gamma V}$ (\ref{L-GV-comp-app}) with identification $\phi'_{2+} = \gamma^4$.
On the other hand, if we evaluate the equation of motion for $\phi'_{2+}$, 
we obtain a trivial equation $0 = \eps^{mn} \del_m \del_n \vartheta$.
Then we immediately obtain the $\vartheta$ sector in the original Lagrangian (\ref{L-Theta-comp}) (or the same form as in (\ref{L-Theta-comp-app})).
Indeed, we have determined the relative coefficients in (\ref{kin-sigmaY2}) in order to realize this structure.
The $(r^3,\phi'_{1-})$ sector also has the same structure, though the interaction term prevents us from obtaining the explicit form of the dual Lagrangian.

Hence we conclude that, in principle,
the Lagrangian $\Scr{L}_{\wt{L}V}$ (\ref{L-tLV-comp4}) generates both the original system of $(r^3, \vartheta)$ and its dualized systems of 
$(r^3, \phi'_{2+})$, $(\phi'_{1-}, \vartheta)$, and $(\phi'_{1-}, \phi'_{2+})$.

\subsubsection*{Lagrangian $\Scr{L}_{LC}$} 

Next, we analyze the Lagrangian $\Scr{L}_{LC}$ (\ref{L-LC-comp}) dualized from the original one $\Scr{L}_{\Psi}$ (\ref{L-PsiC-comp}).
In the same way as (\ref{re-sigma12}), we introduce the following combinations:
\begin{alignat}{2}
\sigma_{1\pm}
\ &\equiv \ 
\sigma_{Y,1} \pm \sigma_{Z,1}
\, , &\ls
\sigma_{2\pm}
\ &\equiv \ 
\sigma_{Y,2} \pm \sigma_{Z,2}
\, . \label{re-phi12}
\end{alignat}
Substituting this and the relations among the auxiliary  fields $(F_X, \wt{G}_Y, \wt{G}_Z)$ into (\ref{L-LC-comp}), we obtain
\begin{align}
\Scr{L}_{LC}
\ &= \ 
- \frac{g^2}{4} (\del_m \sigma_{1+})^2
- \frac{g^2}{4} (\del_m \sigma_{1-})^2
+ \frac{g^2}{2} (\del_m \phi_{X,1})^2
\nn \\
\ & \ \ \ \ 
- r^1 (D_c + \ol{D}{}_c)
+ \frac{g^2}{2} \, \del_+ \del_- \big( \phi_{X,1} - \sigma_{1+} \big) (\phi_{c} + \ol{\phi}{}_{c})
+ g^2 (A_{c=} - \ol{A}{}_{c=}) (B_{c\+} - \ol{B}{}_{c\+})
\nn \\
\ & \ \ \ \ 
- \frac{\I g^2}{2} \, \del_+ \big( \phi_{X,1} + \sigma_{1-} \big) (A_{c=} - \ol{A}{}_{c=})
- \frac{\I g^2}{2} \, \del_- \big( \phi_{X,1} - \sigma_{1-} \big) (B_{c\+} - \ol{B}{}_{c\+})
\nn \\
\ & \ \ \ \
- \frac{g^2}{4} (\del_m \sigma_{2+})^2
- \frac{g^2}{4} (\del_m \sigma_{2-})^2
+ \frac{g^2}{2} (\del_m \phi_{X,2})^2
\nn \\
\ & \ \ \ \ 
- \I r^2 (D_c - \ol{D}{}_c)
- \frac{\I g^2}{2} \, \del_+ \del_- \big( \phi_{X,2} - \sigma_{2-} \big) (\phi_{c} - \ol{\phi}{}_{c})
- g^2 (A_{c=} + \ol{A}{}_{c=}) (B_{c\+} + \ol{B}{}_{c\+})
\nn \\
\ & \ \ \ \ 
- \frac{g^2}{2} \,
\del_+ \big( \phi_{X,2} + \sigma_{2+} \big) (A_{c=} + \ol{A}{}_{c=})
- \frac{g^2}{2} \, \del_- \big( \phi_{X,2} - \sigma_{2+} \big) (B_{c\+} + \ol{B}{}_{c\+})
\nn \\
\ & \ \ \ \
- 2 g^2 |M_{c}|^2
\, . \label{L-LC-comp2} 
\end{align}
Analogous to the previous discussion in (\ref{sigmaY12-abc}),
we rewrite the kinetic term of $\phi_{X,1}$ in terms of (\ref{dr1-L}) and the derivative of (\ref{r1-L}).
The kinetic term of $\phi_{X,2}$ is also rewritten by (\ref{dr2-L}) and the derivative of (\ref{r2-L}).
Substituting (\ref{r1-L}) and (\ref{r2-L}) into the interaction terms,
and integrating out $\sigma_{1+}$ and $\sigma_{2-}$ which do not couple to the original fields, 
we finally obtain the following description:
\begin{align}
\Scr{L}_{LC}
\ &= \ 
- \frac{1}{2 g^2} (\del_m r^1)^2
+ \eps^{mn} (\del_m r^1) (\del_n \sigma_{1-}) 
- \sqrt{2} \, g^2 \eps^{mn} (\del_m \phi_{c,1}) (\del_n \sigma_{1-}) 
\nn \\
\ & \ \ \ \ 
- \frac{1}{2 g^2} (\del_m r^2)^2
- \eps^{mn} (\del_m r^2) (\del_n \sigma_{2+}) 
- \sqrt{2} \, g^2 \eps^{mn} (\del_m \phi_{c,2}) (\del_n \sigma_{2+}) 
\nn \\
\ & \ \ \ \ 
- r^1 (D_c + \ol{D}{}_c)
+ \frac{1}{2} (\del_+ \del_- r^1) (\phi_{c} + \ol{\phi}{}_{c})
%
- \I r^2 (D_c - \ol{D}{}_c)
+ \frac{\I}{2} (\del_+ \del_- r^2) (\phi_{c} - \ol{\phi}{}_{c})
\nn \\
\ & \ \ \ \ 
+ \frac{\I}{2} (\del_+ r^1) (A_{c=} - \ol{A}{}_{c=})
+ \frac{\I}{2} (\del_- r^1) (B_{c\+} - \ol{B}{}_{c\+})
\nn \\
\ & \ \ \ \ 
- \frac{1}{2} (\del_+ r^2) (A_{c=} + \ol{A}{}_{c=})
- \frac{1}{2} (\del_- r^2) (B_{c\+} + \ol{B}{}_{c\+})
- 2 g^2 |M_{c}|^2
\, . \label{L-LC-comp3} 
\end{align}
This Lagrangian also contains two features.
One is the original Lagrangian (\ref{L-LC-comp}) if we integrate out 
the dual fields $\sigma_{1-}$ and $\sigma_{2+}$.
The other is the dual Lagrangian when we integrate out the original fields $r^1$ and $r^2$, though the interaction terms prevent us from performing integration.

We conclude that 
the Lagrangian $\Scr{L}_{LC}$ (\ref{L-LC}) and its alternative (\ref{L-XYZC}),
dualized from $\Scr{L}_{\Psi}$ (\ref{L-Psi}) in terms of the complex linear superfield $L$ and its alternative $X + Y + \ol{Z}$, 
is interpreted as a kind of the first order Lagrangian.
This is because the component expression (\ref{L-LC-comp3}) of $\Scr{L}_{LC}$ involves not only the original fields $(r^1,r^2)$ but also their dual fields $(\sigma_{1-}, \sigma_{2+})$.
If we integrate out one of them, we immediately obtain the second order Lagrangian.
Indeed, $\Scr{L}_{LC}$ provides four second order Lagrangians, i.e.,
the original system of $(r^1,r^2)$ and its dual systems of $(r^1,\sigma_{2+})$, $(\sigma_{1-}, r^2)$, and $(\sigma_{1-}, \sigma_{2+})$.

Summarizing the feature of the Lagrangians (\ref{L-tLV-comp4}) and (\ref{L-LC-comp3}) which contain both the original fields and the dual fields,
we refer to them as ``semi-doubled'' Lagrangians.
In the next section, we will apply the dualization technique to the GLSM for five-branes \cite{Tong:2002rq}.

\section{Semi-doubled GLSM for five-branes}
\label{sect:SD-GLSM}

In this section, we investigate the $\N=(4,4)$ GLSM for five-branes background geometry and its T-dualized systems by virtue of the the duality transformation technique discussed in the previous sections.
We notice that the previous works \cite{Tong:2002rq, Harvey:2005ab, Okuyama:2005gx, Kimura:2013fda, Kimura:2015yla} were based on the duality among the irreducible superfields, while the current work is developed in terms of the reducible superfields $(L, \wt{L})$ and their alternatives.
We will refer to a gauge theory applied such the dualization as the ``semi-doubled'' GLSM.
The benefit is that we can perform the duality transformation along directions even without isometry, i.e.,
we can formally dualize any directions of the transverse space of the five-branes in ten-dimensional string theory.
This implies that we can, in principle, formulate a GLSM whose low energy effective theory is described as the string worldsheet sigma model whose target space would be nongeometric.

\subsection{GLSM by superfields}
\label{sect:sdGLSM-SF}

We begin with the $k$-centered version \cite{Okuyama:2005gx} of the $\N=(4,4)$ abelian GLSM for H-monopoles (smeared NS5-branes) \cite{Tong:2002rq}.
Its Lagrangian $\Scr{L}_{\text{H}}$ in the superfield formalism is given by
\bsubeq \label{L-H}
\begin{align}
\Scr{L}_{\text{H}}
\ &= \ 
\Scr{L}_{\text{H}}^{\text{VM}}
+ \Scr{L}_{\text{H}}^{\text{CHM}}
+ \Scr{L}_{\text{H}}^{\text{NHM}}
\, , \\
\Scr{L}_{\text{H}}^{\text{VM}}
\ &= \
\sum_{a=1}^k \int \d^4 \theta \, 
\frac{1}{e_a^2} \Big\{ - |\Sigma_a|^2 + |\Phi_a|^2 \Big\}
\, , \label{L-H-VM} \\
\Scr{L}_{\text{H}}^{\text{CHM}}
\ &= \
\sum_{a=1}^k \int \d^4 \theta \, \Big\{
|Q_a|^2 \e^{+2V_a} + |\wt{Q}_a|^2 \e^{-2V_a}
\Big\}
- \sum_{a=1}^k \Big\{
\sqrt{2} \int \d^2 \theta \, \wt{Q}_a \Phi_a Q_a 
+ \text{(h.c.)} 
\Big\}
\, , \label{L-H-CHM} \\
\Scr{L}_{\text{H}}^{\text{NHM}}
\ &= \
\frac{1}{g^2} \int \d^4 \theta \, \Big\{ |\Psi|^2 - |\Theta|^2 \Big\}
\nn \\
\ & \ \ \ \ 
+ \sum_{a=1}^k \Big\{
\sqrt{2} \int \d^2 \theta \, (s_a - \Psi) \Phi_a 
+ \sqrt{2} \int \d^2 \wt{\theta} \, (t_a - \Theta) \Sigma_a
+ \text{(h.c.)}
\Big\} 
\, . \label{L-H-NHM}
\end{align}
\esubeq
Here $(\Sigma_a, \Phi_a)$ are $k$ sets of the $\N=(4,4)$ abelian vector multiplets whose building blocks are $\N=(2,2)$ twisted chiral superfields $\Sigma_a$ and $\N=(2,2)$ adjoint chiral superfields $\Phi_a$.
Each of them carries the gauge coupling constant $e_a$.
$Q_a$ and $\wt{Q}_a$ are $\N=(2,2)$ chiral superfields charged $\pm 1$ by $a$-th vector multiplet.
They are the constituents of the $\N=(4,4)$ charged hypermultiplets.
$(s_a, t_a)$ are the complexified Fayet-Iliopoulos (FI) parameters.
They are also expressed as $s_a = \frac{1}{\sqrt{2}} (s_a^1 + \I s_a^2)$ and $t_a = \frac{1}{\sqrt{2}} (t_a^3 + \I t_a^4)$ in terms of the real parameters $(s_a^i, t_a^i)$. 
$(\Psi, \Theta)$ is the $\N=(4,4)$ neutral hypermultiplet constructed by an $\N=(2,2)$ neutral chiral superfield $\Psi$ and an $\N=(2,2)$ twisted chiral superfield $\Theta$.

We consider the duality transformations along the superfields $\Psi$ and $\Theta$ in $\Scr{L}_{\text{H}}^{\text{NHM}}$ (\ref{L-H-NHM}).
They will also give rise to the T-duality transformations of the background geometry in the low energy effective theories, as discussed in \cite{Tong:2002rq, Kimura:2013fda}.
Following the previous sections, we easily obtain 
\begin{align}
\Scr{L}^{\text{NHM}}
\ &= \ 
g^2 \int \d^4 \theta \, 
\Big\{ - \Big| L + 2 \sqrt{2} \sum_a C_a \Big|^2
+ \Big| \wt{L} + 2 \sum_a V_a \Big|^2
\Big\}
\nn \\
\ & \ \ \ \ 
+ \sum_a \Big\{
\sqrt{2} \int \d^2 \theta \, s_a \, \Phi_a 
+ \sqrt{2} \int \d^2 \wt{\theta} \, t_a \, \Sigma_a
+ \text{(h.c.)}
\Big\}
+ \sqrt{2} \sum_a \eps^{mn} \del_m (\vartheta A_{n,a})
\nn \\
\ &= \ 
g^2 \int \d^4 \theta \, 
\Big\{ - \Big| X + Y + \ol{Z} + 2 \sqrt{2} \sum_a C_a \Big|^2
+ \Big| X' + Y' + \ol{W}{}' + 2 \sum_a V_a \Big|^2
\Big\}
\nn \\
\ & \ \ \ \ 
+ \sum_a \Big\{
\sqrt{2} \int \d^2 \theta \, s_a \, \Phi_a 
+ \sqrt{2} \int \d^2 \wt{\theta} \, t_a \, \Sigma_a
+ \text{(h.c.)}
\Big\}
+ \sqrt{2} \sum_a \eps^{mn} \del_m (\vartheta A_{n,a})
\, . \label{L-SD-NHM}
\end{align}
The last term in the right-hand side of (\ref{L-SD-NHM}) is the total derivative term which we ignored in section \ref{sect:duality}.
We can also easily deduce the duality relations from (\ref{Psi2XYZC}) and (\ref{Theta2XYWV}) in such a way as 
\bsubeq \label{SF-relations2}
\begin{align}
\frac{1}{g^2} \ol{\Psi}
\ &= \ 
X + Y + \ol{Z} + 2 \sqrt{2} \sum_a C_a
\, , \\
- \frac{1}{g^2} \ol{\Theta}
\ &= \ 
X' + Y' + \ol{W}{}' + 2 \sum_a V_a
\, .
\end{align}
\esubeq 
It turns out that the resulting form (\ref{L-SD-NHM}) is much simpler than that of \cite{Kimura:2013fda, Kimura:2014aja}.
Moreover, the procedure of the dualization is also quite simple and straightforward.
From now on, we regard the sum of the Lagrangians (\ref{L-H-VM}), (\ref{L-H-CHM}) and (\ref{L-SD-NHM}) under the duality relations (\ref{SF-relations2})
as the ``semi-doubled'' GLSM $\Scr{L}_{\text{SDG}}$ in the superfield formalism.

\subsection{GLSM by component fields}
\label{sect:sdGLSM-comp}

It is straightforward to expand the semi-doubled GLSM $\Scr{L}_{\text{SDG}}$ in terms of the component fields, if we write down the expansion of the superfields as follows:
\bsubeq \label{ex-SF2}
\begin{align}
\Sigma_a 
\ &= \ 
\sigma_a 
+ \I \sqrt{2} \, \theta^+ \ol{\lambda}{}_{+,a} 
- \I \sqrt{2} \, \ol{\theta}{}^- \lambda_{-,a}
- \sqrt{2} \, \theta^+ \ol{\theta}{}^- (D_{V,a} - \I F_{01,a})
+ \ldots
\, , \\
\Phi_a
\ &= \ 
\phi_a 
+ \I \sqrt{2} \, \theta^+ \wt{\lambda}_{+,a} 
+ \I \sqrt{2} \, \theta^- \wt{\lambda}_{-,a}
+ 2 \I \, \theta^+ \theta^- D_{\Phi,a}
+ \ldots
\, , \\
Q_a \ &= \ 
q_a 
+ \I \sqrt{2} \, \theta^+ \psi_{+,a} 
+ \I \sqrt{2} \, \theta^- \psi_{-,a}
+ 2 \I \, \theta^+ \theta^- F_a
+ \ldots
\, , \\
\wt{Q}_a \ &= \ 
\wt{q}_a
+ \I \sqrt{2} \, \theta^+ \wt{\psi}_{+,a}
+ \I \sqrt{2} \, \theta^- \wt{\psi}_{-,a}
+ 2 \I \, \theta^+ \theta^- \wt{F}_a
+ \ldots
\, . 
\end{align}
\esubeq
The expansion of $(\Psi, \Theta, X, Y, \ol{Z}, X', Y', \ol{W}{}')$ has already been exhibited in (\ref{ex-SF}).
The prepotential $C_a$ is also expanded as in (\ref{C}).
We note that the definition $\Phi_a = \ol{D}{}_+ \ol{D}{}_- C_a$ gives the relations among their component fields:
\bsubeq \label{Phi2C}
\begin{align}
\phi_a \ &= \ - \I M_{c,a}
\, , \\
D_{\Phi,a} \ &= \ 
- \I D_{c,a} + \half \del_+ A_{c=,a} + \half \del_- B_{c\+,a} + \frac{\I}{2} \del_+ \del_- \phi_{c,a} 
\, , \\
\wt{\lambda}{}_{\pm,a}
\ &= \ 
- \I \Big( \lambda_{c\pm,a} \pm \del_{\pm} \chi_{c\mp,a} \Big)
\, , \\
&\ls
\{ \, F_{c,a} \, , \ \ 
G_{c,a} \, , \ \ 
N_{c,a} \, , \ \ 
\psi_{c\pm,a} \, , \ \ 
\zeta_{c\pm,a} \, \} \, : \ \ \ 
\text{(no relations)}
\, . 
\end{align}
\esubeq
Substituting the above into the Lagrangian $\Scr{L}_{\text{SDG}}$,
we obtain the explicit form with the component fields.
Since we have already known that there are many redundant fields under 
the duality relations (\ref{SF-relations2}), 
we eliminate them in the same manner as in the last section.
At the end, we obtain the semi-doubled GLSM in terms of the component fields as follows:
\begin{align}
\Scr{L}_{\text{SDG}}
\ &= \ 
\sum_a \frac{1}{e_a^2} \Big\{
\half (F_{01,a})^2 
- |\del_m \sigma_a|^2
- |\del_m M_{c,a}|^2
\Big\}
- \sum_a \Big\{ |D_m q_a|^2 + |D_m \wt{q}_a|^2 \Big\}
\nn \\
\ & \ \ \ \ 
- \frac{1}{2 g^2} \Big\{ (\del_m r^1)^2 + (\del_m r^2)^2 
+ (\del_m r^3)^2 + (\del_m \vartheta)^2 \Big\}
\nn \\
\ & \ \ \ \ 
+ \eps^{mn} (\del_m r^1) (\del_n \sigma_{1-}) 
- \eps^{mn} (\del_m r^2) (\del_n \sigma_{2+}) 
- \eps^{mn} (\del_m r^3) (\del_n \phi'_{1-}) 
+ \eps^{mn} (\del_m \vartheta) (D_n \phi'_{2+}) 
\nn \\
\ & \ \ \ \ 
+ \sqrt{2} \sum_a \eps^{mn} \del_m \big( (\vartheta - t_a^4) A_{n,a} \big)
+ \sqrt{2} \sum_a \eps^{mn} (\del_m \vartheta) A_{n,a} 
\nn \\
\ & \ \ \ \ 
- 2 g^2 \sum_{a,b} \big( \sigma_a \ol{\sigma}{}_b + M_{c,a} \ol{M}{}_{c,b} \big)
- 2 \sum_a \big( |\sigma_a|^2 + |M_{c,a}|^2 \big) \big( |q_a|^2 + |\wt{q}_a|^2 \big)
\nn \\
\ & \ \ \ \ 
- \sum_a \frac{e_a^2}{2} \Big\{ |q_a|^2 - |\wt{q}_a|^2 - \sqrt{2} \, (r^3 - t_a^3) \Big\}^2
- \sum_a e_a^2 \Big| \sqrt{2} \, q_a \wt{q}_a + \big( (r^1 - s_a^1) + \I (r^2 - s_a^2) \big) \Big|^2
\, . \label{L-SD-comp}
\end{align} 
Here we have introduced the gauge covariant derivatives 
$D_m q_a = \del_m q_a - \I A_{m,a} q_a$,
$D_m \wt{q}_a = \del_m \wt{q}_a + \I A_{m,a} \wt{q}_a$,
and 
$D_m \phi'_{2+} = \del_m \phi'_{2+} - 2 \sqrt{2} \sum_a A_{m,a}$.
To make the description simpler, 
we have already integrated out the auxiliary fields\footnote{The explicit evaluation of the auxiliary fields can be seen in \cite{Kimura:2015yla}.}.
We remark that the semi-doubled Lagrangian (\ref{L-SD-comp}) involves 
both the original fields $(r^1,r^2,r^3,\vartheta)$ and the dual fields $(\sigma_{1-}, \sigma_{2+}, \phi'_{1-}, \phi'_{2+})$.
Selecting fields which would be integrated out from this semi-doubled model in an appropriate way, we can obtain various kind of GLSMs for various five-branes.
Here we summarize them in Table \ref{table:sdGLSM}:
\begin{table}[h]
\begin{center}
\slb{.8}{\renewcommand{\arraystretch}{1.3}
\begin{tabular}{c|cccc} \hline
semi-doubled GLSM (\ref{L-SD-comp}) & H-monopoles & KK-monopoles & $5^2_2$-brane & $5^3_2$-brane
\\ \hline\hline
dynamical & $(r^1,r^2,r^3,\vartheta)$ & $(r^1,r^2,r^3,\phi'_{2+})$ & $(r^1,r^2, \sigma_{2+}, r^3, \phi'_{2+})$ & $(r^1, \sigma_{1-}, r^2, \sigma_{2+}, r^3, \phi'_{2+})$
\\
integrated-out & $(\sigma_{1-},\sigma_{2+},\phi'_{1-},\phi'_{2+})$ & $(\sigma_{1-},\sigma_{2+},\phi'_{1-},\vartheta)$ & $(\sigma_{1-}, \phi'_{1-}, \vartheta)$ & $(\phi'_{1-}, \vartheta)$
\\
co-existing & -- & -- & $(r^2,\sigma_{2+})$ & $(r^1,\sigma_{1-})$, $(r^2, \sigma_{2+})$ 
\\ \hline
\end{tabular}
}
\end{center}
\caption{Various GLSMs from the semi-doubled GLSM $\Scr{L}_{\text{SDG}}$ (\ref{L-SD-comp}).}
\label{table:sdGLSM}
\end{table}

\noindent
We have several comments on Table \ref{table:sdGLSM}:
In the first row we listed various five-branes.
We start from H-monopoles (smeared NS5-branes).
Performing T-duality along $\vartheta$, KK-monopoles appear.
Taking further T-duality along $r^2$, 
the background of the KK-monopoles is transformed to that of an exotic $5^2_2$-brane which is locally geometric but globally nongeometric \cite{deBoer:2010ud}.
Dualizing one more direction $r^1$, an exotic $5^3_2$-brane will appear.
This five-brane background would be even locally nongeometric.
Hence, in the second row the dynamical fields in each GLSM are listed,
whilst in the third row the fields integrated out are described. 
The backgrounds of the H-monopoles and the KK-monopoles have an isometry along $\vartheta$ and $\phi'_{2+}$, respectively.
However, the backgrounds of the exotic branes have no isometry along $r^2$ and $r^1$ directions.
Then, in the GLSM level, both the original field and the dual field simultaneously exist.
This is exhibited in the fourth row.

\subsection{Various GLSMs}
\label{sect:GLSMs}

In this subsection, 
we derive the GLSMs for various five-branes from the semi-doubled GLSM (\ref{L-SD-comp}) as listed in Table \ref{table:sdGLSM}.
Although the computation itself is quite straightforward, 
the result will play an instructive role in considering the low energy effective theories in the next section.

\subsubsection*{H-monopoles}

First, we derive the GLSM for H-monopoles.
As seen in Table \ref{table:sdGLSM}, this is a gauge theory of dynamical fields $(r^1,r^2,r^3, \vartheta)$.
We integrate out the other fields $(\sigma_{1-}, \sigma_{2+}, \phi'_{1-}, \phi'_{2+})$ from the semi-doubled GLSM (\ref{L-SD-comp}).
Their equations of motion provide trivial equations as
\bsubeq \label{EOM-SDG-H}
\begin{alignat}{2}
0 \ &= \ 
\del^m \Big\{
+ \eps_{mn} \del^n r^1
\Big\}
\, , &\ls
0 \ &= \
\del^m \Big\{
- \eps_{mn} \del^n r^2
\Big\}
\, , \\
0 \ &= \ 
\del^m \Big\{
- \eps_{mn} \del^n r^3
\Big\}
\, , &\ls
0 \ &= \  
\del^m \Big\{
+ \eps_{mn} \del^n \vartheta
\Big\}
\, . 
\end{alignat}
\esubeq
Then we just remove the terms of $(\sigma_{1-}, \sigma_{2+}, \phi'_{1-}, \phi'_{2+})$ from (\ref{L-SD-comp}).
The resulting Lagrangian is 
\begin{align}
\Scr{L}_{\text{H}}
\ &= \ 
\sum_a \frac{1}{e_a^2} \Big\{
\half (F_{01,a})^2 
- |\del_m \sigma_a|^2
- |\del_m M_{c,a}|^2
\Big\}
- \sum_a \Big\{ |D_m q_a|^2 + |D_m \wt{q}_a|^2 \Big\}
\nn \\
\ & \ \ \ \ 
- \frac{1}{2 g^2} \Big\{ (\del_m r^1)^2 + (\del_m r^2)^2 
+ (\del_m r^3)^2 + (\del_m \vartheta)^2 \Big\}
\nn \\
\ & \ \ \ \ 
+ \sqrt{2} \sum_a \eps^{mn} \del_m \big( (\vartheta - t_a^4) A_{n,a} \big)
- \sqrt{2} \sum_a \eps^{mn} (\del_m \vartheta) A_{n,a} 
\nn \\
\ & \ \ \ \ 
- 2 g^2 \sum_{a,b} \big( \sigma_a \ol{\sigma}{}_b + M_{c,a} \ol{M}{}_{c,b} \big)
- 2 \sum_a \big( |\sigma_a|^2 + |M_{c,a}|^2 \big) \big( |q_a|^2 + |\wt{q}_a|^2 \big)
\nn \\
\ & \ \ \ \ 
- \sum_a \frac{e_a^2}{2} \Big\{ |q_a|^2 - |\wt{q}_a|^2 - \sqrt{2} \, (r^3 - t_a^3) \Big\}^2
- \sum_a e_a^2 \Big| \sqrt{2} \, q_a \wt{q}_a + \big( (r^1 - s_a^1) + \I (r^2 - s_a^2) \big) \Big|^2
\, . \label{L-SD-comp-Hmonopole}
\end{align} 
Up to the total derivative term, this is nothing but the GLSM for H-monopoles discussed in \cite{Tong:2002rq, Harvey:2005ab, Okuyama:2005gx}.

\subsubsection*{KK-monopoles}

Second, we consider the GLSM for KK-monopoles from (\ref{L-SD-comp}).
Table \ref{table:sdGLSM} indicates that we should integrate out $(\sigma_{1-}, \sigma_{2+}, \phi'_{1-}, \vartheta)$.
This is also straightforward.
Evaluate the equations of motion for them:
\bsubeq \label{EOM-SDG-KK}
\begin{alignat}{2}
0 \ &= \ 
\del^m \Big\{
+ \eps_{mn} \del^n r^1
\Big\}
\, , &\ls
0 \ &= \
\del^m \Big\{
- \eps_{mn} \del^n r^2
\Big\}
\, , \\
0 \ &= \ 
\del^m \Big\{
- \eps_{mn} \del^n r^3
\Big\}
\, , &\ls
0 \ &= \  
\del^m \Big\{
\frac{1}{g^2} \del_m \vartheta
- \eps_{mn} D^n \phi'_{2+}
- \sqrt{2} \sum_a \eps_{mn} A^n_a
\Big\}
\, . 
\end{alignat}
\esubeq
Plugging this into (\ref{L-SD-comp}), we obtain 
\begin{align}
\Scr{L}_{\text{KK}}
\ &= \ 
\sum_a \frac{1}{e_a^2} \Big\{
\half (F_{01,a})^2 
- |\del_m \sigma_a|^2
- |\del_m M_{c,a}|^2
\Big\}
- \sum_a \Big\{ |D_m q_a|^2 + |D_m \wt{q}_a|^2 \Big\}
\nn \\
\ & \ \ \ \ 
- \frac{1}{2 g^2} \Big\{ (\del_m r^1)^2 + (\del_m r^2)^2 + (\del_m r^3)^2 \Big\}
- \frac{g^2}{2} (D_m \gamma^4)^2
+ \sqrt{2} \sum_a \eps^{mn} \del_m \big( (\vartheta - t_a^4) A_{n,a} \big)
\nn \\
\ & \ \ \ \ 
- 2 g^2 \sum_{a,b} \big( \sigma_a \ol{\sigma}{}_b + M_{c,a} \ol{M}{}_{c,b} \big)
- 2 \sum_a \big( |\sigma_a|^2 + |M_{c,a}|^2 \big) \big( |q_a|^2 + |\wt{q}_a|^2 \big)
\nn \\
\ & \ \ \ \ 
- \sum_a \frac{e_a^2}{2} \Big\{ |q_a|^2 - |\wt{q}_a|^2 - \sqrt{2} \, (r^3 - t_a^3) \Big\}^2
- \sum_a e_a^2 \Big| \sqrt{2} \, q_a \wt{q}_a + \big( (r^1 - s_a^1) + \I (r^2 - s_a^2) \big) \Big|^2
\, . \label{L-SD-comp-KK}
\end{align} 
Here we rewrote $\phi'_{2+}$ as $\gamma^4$ and introduced its gauge covariant derivative as
\begin{align}
D_m \phi'_{2+} + \sqrt{2} \sum_a A_{m,a}
\ &= \ 
\del_m \gamma^4 - \sqrt{2} \sum_a A_{m,a}
\ \equiv \ 
D_m \gamma^4
\, . \label{Dgamma4}
\end{align}
This is the correct form of the GLSM for KK-monopoles \cite{Tong:2002rq, Harvey:2005ab, Okuyama:2005gx}.

\subsubsection*{$5^2_2$-brane}

Let us derive the GLSM for an exotic $5^2_2$-brane discussed in \cite{Kimura:2013fda, Kimura:2015yla}.
As discussed before, the $5^2_2$-brane is generated by the smearing along one direction without isometry in the background geometry of the KK-monopoles.
Suppose we perform the duality transformation of the original field $r^2$ and obtain the system of its dual field $\sigma_{2+}$.
However, they do not have isometry.
Then we construct the GLSM for the $5^2_2$-brane by integrating out only $(\sigma_{1-}, \phi'_{1-}, \vartheta)$, while both $r^2$ and $\sigma_{2+}$ are not integrated out:
\bsubeq \label{EOM-SDG-522}
\begin{align}
0 \ &= \ 
\del^m \Big\{
+ \eps_{mn} \del^n r^1
\Big\}
\, , \ls
0 \ = \ 
\del^m \Big\{
- \eps_{mn} \del^n r^3
\Big\}
\, , \\
0 \ &= \  
\del^m \Big\{
\frac{1}{g^2} \del_m \vartheta
- \eps_{mn} D^n \phi'_{2+}
- \sqrt{2} \sum_a \eps_{mn} A^n_a
\Big\}
\, . 
\end{align}
\esubeq
Substituting them into the semi-doubled GLSM (\ref{L-SD-comp}), 
we find 
\begin{align}
\Scr{L}_{5^2_2}
\ &= \ 
\sum_a \frac{1}{e_a^2} \Big\{
\half (F_{01,a})^2 
- |\del_m \sigma_a|^2
- |\del_m M_{c,a}|^2
\Big\}
- \sum_a \Big\{ |D_m q_a|^2 + |D_m \wt{q}_a|^2 \Big\}
\nn \\
\ & \ \ \ \ 
- \frac{1}{2 g^2} \Big\{ (\del_m r^1)^2 + (\del_m r^2)^2 
+ (\del_m r^3)^2 \Big\}
- \frac{g^2}{2} (D_m \gamma^4)^2
\nn \\
\ & \ \ \ \ 
- \eps^{mn} (\del_m r^2) (\del_n \sigma_{2+}) 
+ \sqrt{2} \sum_a \eps^{mn} \del_m \big( (\vartheta - t_a^4) A_{n,a} \big)
\nn \\
\ & \ \ \ \ 
- 2 g^2 \sum_{a,b} \big( \sigma_a \ol{\sigma}{}_b + M_{c,a} \ol{M}{}_{c,b} \big)
- 2 \sum_a \big( |\sigma_a|^2 + |M_{c,a}|^2 \big) \big( |q_a|^2 + |\wt{q}_a|^2 \big)
\nn \\
\ & \ \ \ \ 
- \sum_a \frac{e_a^2}{2} \Big\{ |q_a|^2 - |\wt{q}_a|^2 - \sqrt{2} \, (r^3 - t_a^3) \Big\}^2
- \sum_a e_a^2 \Big| \sqrt{2} \, q_a \wt{q}_a + \big( (r^1 - s_a^1) + \I (r^2 - s_a^2) \big) \Big|^2
\, . \label{L-SD-comp-522}
\end{align} 
Here we replaced $(\sigma_{2+}, \phi'_{2+})$ to $(-y^2, \gamma^4)$ and introduced $D_m \gamma^4$ (\ref{Dgamma4}).
It turns out that this is still the ``semi-doubled'' model.
Moreover, this is nothing but the GLSM for the $5^2_2$-brane proposed in \cite{Kimura:2013fda, Kimura:2015yla}.

\subsubsection*{$5^3_2$-brane}

Finally we argue the GLSM for an exotic $5^3_2$-brane.
This five-brane is obtained by T-duality along the three directions of the transverse space of the H-monopoles.
This background is regarded as a locally nongeometric background.
Then there are explicitly no descriptions as a conventional geometry.
Fortunately, however, 
we can formally describe its GLSM where the original and dual fields coexist. 
This is the same strategy as in doubled sigma model and double field theory
\cite{Siegel:1993xq, Siegel:1993th, Siegel:1993bj, Hull:2004in, Hull:2006va, Dall'Agata:2008qz, Albertsson:2008gq, Hull:2009sg, Hull:2009mi, Hohm:2010jy, Zwiebach:2011rg, Aldazabal:2013sca, Berman:2013eva}, 
$\beta$-supergravity and related geometry
\cite{Andriot:2013xca, Andriot:2014uda, Asakawa:2014kua, Asakawa:2015jza}, 
and so forth.

Following Table \ref{table:sdGLSM}, we integrate out only $(\phi'_{1-}, \vartheta)$ from the semi-doubled GLSM (\ref{L-SD-comp}):
\begin{align}
0 \ &= \ 
\del^m \Big\{
- \eps_{mn} \del^n r^3
\Big\}
\, , \ls
0 \ = \  
\del^m \Big\{
\frac{1}{g^2} \del_m \vartheta
- \eps_{mn} D^n \phi'_{2+}
- \sqrt{2} \sum_a \eps_{mn} A^n_a
\Big\}
\, . \label{EOM-SDG-532}
\end{align}
Then the Lagrangian is reduced to the following form:
\begin{align}
\Scr{L}_{5^3_2}
\ &= \ 
\sum_a \frac{1}{e_a^2} \Big\{
\half (F_{01,a})^2 
- |\del_m \sigma_a|^2
- |\del_m M_{c,a}|^2
\Big\}
- \sum_a \Big\{ |D_m q_a|^2 + |D_m \wt{q}_a|^2 \Big\}
\nn \\
\ & \ \ \ \ 
- \frac{1}{2 g^2} \Big\{ (\del_m r^1)^2 + (\del_m r^2)^2 
+ (\del_m r^3)^2 \Big\}
- \frac{g^2}{2} (D_m \gamma^4)^2
\nn \\
\ & \ \ \ \ 
+ \eps^{mn} (\del_m r^1) (\del_n \sigma_{1-}) 
- \eps^{mn} (\del_m r^2) (\del_n \sigma_{2+}) 
+ \sqrt{2} \sum_a \eps^{mn} \del_m \big( (\vartheta - t_a^4) A_{n,a} \big)
\nn \\
\ & \ \ \ \ 
- 2 g^2 \sum_{a,b} \big( \sigma_a \ol{\sigma}{}_b + M_{c,a} \ol{M}{}_{c,b} \big)
- 2 \sum_a \big( |\sigma_a|^2 + |M_{c,a}|^2 \big) \big( |q_a|^2 + |\wt{q}_a|^2 \big)
\nn \\
\ & \ \ \ \ 
- \sum_a \frac{e_a^2}{2} \Big\{ |q_a|^2 - |\wt{q}_a|^2 - \sqrt{2} \, (r^3 - t_a^3) \Big\}^2
- \sum_a e_a^2 \Big| \sqrt{2} \, q_a \wt{q}_a + \big( (r^1 - s_a^1) + \I (r^2 - s_a^2) \big) \Big|^2
\, . \label{L-SD-comp-532}
\end{align} 
This is still a ``semi-doubled'' GLSM because the dual fields $(\sigma_{1-}, \sigma_{2+})$ have no canonical kinetic terms, while they contribute to the system.
In the next section we will argue how this model gives the nongeometric structure.

We summarize this section.
We started from the semi-doubled GLSM for five-branes (\ref{L-SD-comp}).
Integrating out certain fields, we obtained the conventional GLSMs for various five-branes.
All of them, expect for (\ref{L-SD-comp-532}), are the models which have already been obtained in previous works.
The procedure of integration is quite simple.
In addition, we proposed the GLSM for the exotic $5^3_2$-brane, although we have no ideas how to justify it in the current stage.

\section{Semi-doubled NLSM for five-branes}
\label{sect:SD-NLSM}

In this section, we investigate the low energy effective theory of the semi-doubled GLSM $\Scr{L}_{\text{SDG}}$ (\ref{L-SD-comp}) discussed in the last section.
The low energy effective theory is given as a NLSM which still involves the original and dual fields.
Hence we will refer to this as the ``semi-doubled'' NLSM.
Integrating out a certain set of fields, we will obtain conventional NLSMs whose target spaces are five-brane backgrounds.
Independently, we also briefly mention the low energy effective theories of the various GLSMs for five-branes obtained in the previous section.
They will correspond to the ones derived from the semi-doubled NLSM.

\subsection{Low energy limit of the semi-doubled GLSM}
\label{sect:IR-sdNLSM}

We explore the supersymmetric low energy effective theory.
The Lagrangian (\ref{L-SD-comp}) has the potential terms which vanish on the supersymmetric vacua:
\bsubeq \label{vacua-eqs}
\begin{alignat}{2}
0 \ &= \ 
\sum_{a,b} \big( \sigma_a \ol{\sigma}{}_b + M_{c,a} \ol{M}{}_{c,b} \big)
\, , &\ls
0 \ &= \ 
\big( |\sigma_a|^2 + |M_{c,a}|^2 \big) \big( |q_a|^2 + |\wt{q}_a|^2 \big)
\, , \label{vacua-VM} \\
0 \ &=  \
|q_a|^2 - |\wt{q}_a|^2 - \sqrt{2} \, (r^3 - t_a^3) 
\, , &\ls
0 \ &= \ \sqrt{2} \, q_a \wt{q}_a + \big( (r^1 - s_a^1) + \I (r^2 - s_a^2) \big)
\, . \label{vacua-qtq}
\end{alignat}
\esubeq
We focus only on the Higgs phase in which all of the scalar fields of the vector multiplets vanish.
Then the first two equations (\ref{vacua-VM}) are trivial.
Furthermore, we can solve the second two equations (\ref{vacua-qtq}) with respect to the complex scalar fields $(q_a, \wt{q}_a)$ \cite{Tong:2002rq, Harvey:2005ab}:
\bsubeq \label{sol-vacua-qtq}
\begin{gather}
q_a \ = \ 
\frac{\I}{2^{1/4}} \, \e^{+\I \alpha_a} 
\sqrt{R_a + (r^3 - t_a^3)}
\, , \ls
\wt{q}_a \ = \ 
\frac{\I}{2^{1/4}} \, \e^{-\I \alpha_a}
\frac{(r^1 - s_a^1) + \I (r^2 - s_a^2)}{\sqrt{R_a + (r^3 - t_a^3)}}
\, , \\
(R_a)^2 \ \equiv \ 
(r^1 - s_a^1)^2 + (r^2 - s_a^2)^2 + (r^3 - t_a^3)^2
\, . 
\end{gather}
\esubeq
Here $\alpha_a$ is the gauge parameter.
Plugging this solution into each kinetic term of $(q_a, \wt{q}_a)$, 
we obtain the following form:
\bsubeq \label{DqDtq}
\begin{align}
- \Big\{ |D_m q_a|^2 + |D_m \wt{q}_a|^2 \Big\}
\ &= \ 
- \frac{1}{2 \sqrt{2} \, R_a} \Big\{
(\del_m r^1)^2
+ (\del_m r^2)^2
+ (\del_m r^3)^2
\Big\}
\nn \\
\ & \ \ \ \ 
- \sqrt{2} \, R_a \Big(
\del_m \alpha_a 
- A_{m,a}
+ \frac{1}{\sqrt{2}} \, \Omega_{i,a} \, \del_m r^i
\Big)^2
\, , \\
\Omega_{i,a} \, \del_m r^i
\ &\equiv \ 
\frac{- (r^1 - s_a^1) \del_m r^2 + (r^2 - s_a^2) \del_m r^1}{\sqrt{2} \, R_a (R_a + (r^3 - t_a^3))}
\, . \label{Omega-522S}
\end{align}
\esubeq
For later convenience, we refer to $\Omega_{i,a}$ as the KK-vector.
The KK-vector will play a significant role in analyzing the target space structure of the low energy effective theory.
We note that the third component $\Omega_{3,a}$ is trivial\footnote{The triviality is just an artifact of the explicit construction of $\N=(4,4)$ theory in terms of $\N=(2,2)$ supermultiplets. Indeed there exists $SU(2)_R$ symmetry in this system. Under this R-symmetry the vectors $r^i$ and $\Omega_{i,a}$ behave as the triplets.}.
Since we have solved the equations of the supersymmetric vacua (\ref{vacua-eqs}), 
the Lagrangian is reduced to
\begin{align}
\Scr{L}_{\text{SDG}}
\ &= \ 
\sum_a \frac{1}{2 e_a^2} (F_{01,a})^2 
- \sum_a \sqrt{2} \, R_a \Big(
\del_m \alpha_a 
- A_{m,a}
+ \frac{1}{\sqrt{2}} \, \Omega_{i,a} \, \del_m r^i
\Big)^2
+ \eps^{mn} (\del_m \vartheta) (D_n \gamma^4) 
\nn \\
\ & \ \ \ \ 
- \frac{H}{2} \Big\{ (\del_m r^1)^2 + (\del_m r^2)^2 + (\del_m r^3)^2 \Big\}
- \frac{1}{2 g^2} (\del_m \vartheta)^2 
\nn \\
\ & \ \ \ \ 
+ \eps^{mn} (\del_m r^1) (\del_n \sigma_{1-}) 
- \eps^{mn} (\del_m r^2) (\del_n \sigma_{2+}) 
- \eps^{mn} (\del_m r^3) (\del_n \phi'_{1-}) 
\nn \\
\ & \ \ \ \ 
+ \sqrt{2} \sum_a \eps^{mn} \del_m \big( (\vartheta - t_a^4) A_{n,a} \big)
\, . \label{L-SD-comp2}
\end{align}
Here we introduced a harmonic function $H$ 
whose divergence is given as a rotation of the KK-vector:
\bsubeq
\begin{align}
H \ &\equiv \ 
\frac{1}{g^2} + \sum_a \frac{1}{\sqrt{2} \, R_a}
\, , \label{cod3-H-522S} \\
\nabla_i H \ &= \ (\nabla \times \vec{\Omega})_i
\, , \ls
\Omega_i \ \equiv \ 
\sum_a \Omega_{i,a}
\, . \label{monopole-eq}
\end{align}
\esubeq
The equation (\ref{monopole-eq}) is interpreted as the Dirac monopole equation.

Next, we take the low energy limit of the model (\ref{L-SD-comp2}).
This is controlled by the infinity limit of the gauge coupling constants $e_a \to \infty$ because they are of mass dimension one.
Since the dynamics of the gauge fields $A_{m,a}$ are frozen in this limit,
we integrate them out.
The solution of the equation of motion for each gauge field is 
\begin{align}
A_{m,a} 
\ &= \  
\del_m \alpha_a 
+ \frac{1}{\sqrt{2}} \, \Omega_{i,a} \, \del_m r^i
+ \frac{1}{2 R_a} \eps_{mn} \, \del^n \vartheta
\, . \label{A-sol}
\end{align}
We have a comment that we can quite easily obtain the solution compared with the case of the GLSM for KK-monopoles demonstrated in \cite{Okuyama:2005gx}.
Plugging this into (\ref{L-SD-comp2}) in the low energy limit,
we obtain the following form:
\begin{align}
\Scr{L}_{\text{SDN}}
\ &= \ 
- \frac{H}{2} \Big\{ (\del_m r^1)^2 + (\del_m r^2)^2 + (\del_m r^3)^2 + (\del_m \vartheta)^2 \Big\}
\nn \\
\ & \ \ \ \ 
+ \eps^{mn} (\del_m r^1) (\del_n \sigma_{1-}) 
- \eps^{mn} (\del_m r^2) (\del_n \sigma_{2+}) 
- \eps^{mn} (\del_m r^3) (\del_n \phi'_{1-}) 
\nn \\
\ & \ \ \ \ 
+ \eps^{mn} (\del_m \vartheta) \Big( \del_n \wt{\vartheta} - \Omega_i \, \del_n r^i \Big) 
+ \sqrt{2} \sum_a \eps^{mn} \del_m \big( (\vartheta - t_a^4) \mr{A}_{n,a} \big)
\, . \label{L-SDN-comp}
\end{align}
Here we introduced the gauge invariant field $\wt{\vartheta} \equiv \gamma^4 - \sqrt{2} \sum_a \alpha_a$.
This is genuinely the dual field of $\vartheta$.
In the topological term, $\mr{A}_{n,a}$ indicates that we substituted the solution (\ref{A-sol}) into this term.
This is the low energy effective theory of the semi-doubled GLSM.
We refer to this as the ``semi-doubled'' NLSM because both the original fields $(r^1,r^2,r^3,\vartheta)$ and the dual fields $(\sigma_{1-}, \sigma_{2+}, \phi'_{1-}, \wt{\vartheta})$ are involved,
though the latter contributes to the system only topologically.
We can derive various NLSMs whose target spaces are backgrounds of five-branes,
if we integrate out a certain set of original and/or dual fields as discussed at the GLSM level.

\subsection{Low energy limit of various GLSMs}
\label{sect:IR-NLSMs}

Here we integrate out a certain set of fields from the semi-doubled NLSM $\Scr{L}_{\text{SDN}}$ (\ref{L-SDN-comp}) and obtain the various NLSMs for five-branes.
First we summarize the configurations which we analyze in Table \ref{table:sdNLSM}:
\begin{table}[h]
\begin{center}
\slb{.8}{\renewcommand{\arraystretch}{1.3}
\begin{tabular}{c|cccc} \hline
semi-doubled NLSM (\ref{L-SDN-comp}) & H-monopoles & KK-monopoles & $5^2_2$-brane & ``$5^3_2$-brane''
\\ \hline\hline
dynamical & $(r^1,r^2,r^3,\vartheta)$ & $(r^1,r^2,r^3,\wt{\vartheta})$ &
$(r^1,\sigma_{2+}, r^3, \wt{\vartheta})$ & $(\sigma_{1-}, \sigma_{2+}, r^3, \wt{\vartheta})$
\\
smearing & -- & -- & $r^2$ & $(r^1,r^2)$
\\
integrated-out & $(\sigma_{1-}, \sigma_{2+}, \phi'_{1-}, \wt{\vartheta})$
& $(\sigma_{1-}, \sigma_{2+}, \phi'_{1-}, \vartheta)$
& $(\sigma_{1-}, r^2, \phi'_{1-}, \vartheta)$
& $(r^1, r^2, \phi'_{1-}, \vartheta)$
\\ \hline
\end{tabular}
}
\end{center}
\caption{Various NLSMs which will be derived from the semi-doubled NLSM (\ref{L-SDN-comp}).}
\label{table:sdNLSM}
\end{table}

\noindent
We have comments on Table \ref{table:sdNLSM}.
The second row exhibits the dynamical fields which govern the NLSM.
We prepare the third row because we have to make isometry along certain directions in order to obtain the backgrounds of the exotic five-branes.
The fourth row describes the fields which should be integrated out {\it after} smearing the fields in the second row.
In the fifth column in the first row, we express the name of the background with double-quotation marks because we will not be able to obtain the conventional description of this background.
We will discuss this issue later.

From now on we derive the various NLSMs from the semi-doubled NLSM.
In each model we briefly mention the low energy limit of the corresponding GLSMs which we obtained in section \ref{sect:GLSMs}.

\subsubsection*{H-monopoles}

Following Table \ref{table:sdNLSM}, 
we integrate out the dual fields $(\sigma_{1-}, \sigma_{2+}, \phi'_{1-}, \wt{\vartheta})$ in the semi-doubled NLSM (\ref{L-SDN-comp}).
It turns out that their field equations are trivial as seen in (\ref{EOM-SDG-H}):
\bsubeq \label{EOM-red-H-SD}
\begin{alignat}{2}
0 \ &= \ 
\del^m \Big\{
+ \eps_{mn} \del^n r^1
\Big\}
\, , &\ls 
0 \ &= \ 
\del^m \Big\{
- \eps_{mn} \del^n r^2
\Big\}
\, , \\
0 \ &= \ 
\del^m \Big\{
- \eps_{mn} r^3
\Big\}
\, , &\ls
0 \ &= \ 
\del^m \Big\{
+ \eps_{mn} \del^n \vartheta
\Big\}
\, .
\end{alignat}
\esubeq
We can simply remove the terms containing the dual fields.
Then we obtain
\begin{align}
\Scr{L}_{\text{H}}
\ &= \ 
- \frac{H}{2} \Big\{ (\del_m r^1)^2 + (\del_m r^2)^2 + (\del_m r^3)^2 + (\del_m \vartheta)^2 \Big\}
+ \eps^{mn} \, \Omega_i \, (\del_m r^i) (\del_n \vartheta) 
\nn \\
\ & \ \ \ \ 
+ \sqrt{2} \sum_a \eps^{mn} \del_m \big( (\vartheta - t_a^4) \mr{A}_{n,a} \big)
\, . \label{L-SD-H}
\end{align}
This is nothing but the NLSM whose target space is the background configuration of the H-monopoles. 
Following the procedure in section \ref{sect:IR-sdNLSM},
this Lagrangian also appears as the low energy limit of the GLSM (\ref{L-SD-comp-Hmonopole}).
Indeed the analysis of the low energy limit of (\ref{L-SD-comp-Hmonopole}) was demonstrated in \cite{Tong:2002rq, Harvey:2005ab, Okuyama:2005gx}.

\subsubsection*{KK-monopoles}

We analyze the NLSM following the third column in Table \ref{table:sdNLSM}.
First, we integrate out the fields $(\sigma_{1-},\sigma_{2+}, \phi'_{1-}, \vartheta)$.
Their field equations are
\bsubeq \label{EOM-red-KK-SD}
\begin{alignat}{2}
0 \ &= \ 
\del^m \Big\{
+ \eps_{mn} \del^n r^1
\Big\}
\, , &\ls
0 \ &= \ 
\del^m \Big\{
- \eps_{mn} \del^n r^2
\Big\}
\, , \\
0 \ &= \ 
\del^m \Big\{
- \eps_{mn} r^3
\Big\}
\, , &\ls
0 \ &= \ 
\del^m \Big\{
H \, \del_m \vartheta
- \eps_{mn} \Big( \del^n \wt{\vartheta} - \Omega_i \, \del^n r^i \Big)
\Big\}
\, .
\end{alignat}
\esubeq
Only the field equation for $\vartheta$ is non-trivial.
Applying them to the semi-doubled NLSM (\ref{L-SDN-comp}),
we obtain the conventional form of the NLSM whose target space is the background geometry of the KK-monopoles \cite{Tong:2002rq, Harvey:2005ab, Okuyama:2005gx}:
\begin{align}
\Scr{L}_{\text{KK}}
\ &= \ 
- \frac{H}{2} \Big\{ (\del_m r^1)^2 + (\del_m r^2)^2 + (\del_m r^3)^2 \Big\}
- \frac{1}{2 H} \Big( \del_m \wt{\vartheta} - \Omega_i \, \del_m r^i \Big)^2
\nn \\
\ & \ \ \ \ 
+ \sqrt{2} \sum_a \eps^{mn} \del_m \big( (\vartheta - t_a^4) \mr{A}_{n,a} \big)
\, . \label{L-SD-KK}
\end{align}
This is also obtained by the low energy limit of the GLSM (\ref{L-SD-comp-KK})
through the the procedure demonstrated in section \ref{sect:IR-sdNLSM}.

\subsubsection*{$5^2_2$-brane}

We analyze the low energy theory in the fourth column in Table \ref{table:sdNLSM}.
Before doing this, we should keep in mind that the functions $(H, \Omega_i)$ depend on the field $r^2$.
In order to construct an isometry along the field $r^2$ (and its dual $\sigma_{2+}$),
we perform the smearing procedure and make $(H, \Omega_i)$ independent of $r^2$ \cite{Sen:1994wr, Blau:1997du, Cherkis:2000cj, Cherkis:2001gm, deBoer:2010ud, Kikuchi:2012za}.

First, for simplicity, we set $(s_a^1, t_a^3) = (0,0)$ for each FI parameter.
Next, we set that the field $r^2$ moves only on a circle of radius ${\cal R}_2$.
In other words, we introduce a periodicity with period $2 \pi {\cal R}_2 \equiv \Delta s_a^2$.
In addition, we set the second FI parameter $s_a^2$ to
\begin{align}
s_a^2 \ &= \ 2 \pi {\cal R}_2 \, a \ \equiv \ x
\, . \label{FI-s2}
\end{align}
In the small limit ${\cal R}_2 \to 0$, 
the period $2 \pi {\cal R}_2$ is also infinitesimally small with $\Delta s_a^2 \to \d x$.
This also implies the large limit $k \to \infty$.
Then we can replace the sum with respect to $a$ in $(H, \Omega_i)$ to the integration as in the following forms:
\bsubeq \label{smearing2-SD}
\begin{align}
(R_{a})^2 \ &= \ 
(r^1)^2 + (r^2 - s_{a}^2)^2 + (r^3)^2
\ = \ 
\varrho^2 + (r^2 - x)^2
\, , \\
r^1 \ &\equiv \ \varrho \, \cos \vartheta_{\varrho}
\, , \ls
r^3 \ \equiv \ \varrho \, \sin \vartheta_{\varrho}
\, , \\
H
\ &= \ 
\frac{1}{g^2} + \lim_{k\to\infty} \sum_{a=1}^{k} \frac{1}{\sqrt{2} \, R_{a}}
\ = \ 
\frac{1}{g^2} + \lim_{L\to \infty} \frac{1}{2 \pi {\cal R}_2} \int_{-L}^L \frac{\d x}{\sqrt{2} \, R_{a}}
\nn \\
\ &= \ 
\frac{1}{g^2} + \sigma'' \log \frac{\Lambda}{\varrho}
\, , \ls
\sigma'' \ \equiv \ \frac{1}{\sqrt{2} \, \pi {\cal R}_2} 
\, , \\
\Omega_1
\ &= \ 
\lim_{k\to\infty} \sum_{a=1}^{k} \Omega_{1,a}
\ = \ 
\frac{1}{2 \pi {\cal R}_2} \lim_{L \to \infty} \int_{-L}^L \d x \, \frac{r^2 - x}{\sqrt{2} \, R_{a} (R_{a} + (r^3 - t_j^3))}
\nn \\
\ &= \ 
0
\, , \\
\Omega_2
\ &= \ 
\lim_{k\to\infty} \sum_{a=1}^{k} \Omega_{2,a}
\ = \ 
- \frac{r^1}{2 \pi {\cal R}_2} \lim_{L\to\infty} \int_{-L}^L 
\frac{\d x}{\sqrt{2} \, R_{a} (R_{a} + r^3)}
\nn \\
\ &= \ 
\sigma'' \vartheta_{\varrho}
+ \text{(divergent part)}
\, . 
\end{align}
\esubeq
Here $L$ is the cut-off and $\Lambda$ is a divergent parameter.
We notice that $\Omega_3$ vanishes from the beginning.
Due to this procedure, the functions $(H, \Omega_i)$ do not depend on $r^2$ any more, though they still satisfy the Dirac monopole equation (\ref{monopole-eq}).
Plugging the finite parts of (\ref{smearing2-SD}) into (\ref{L-SDN-comp}), we have the semi-doubled Lagrangian which do not depend on the non-derivative $r^2$.
Then we integrate out the fields $(\sigma_{1-},r^2,\phi'_{1-}, \vartheta)$ as mentioned in Table \ref{table:sdNLSM}.
The field equations of $(\sigma_{1-},\phi'_{1-})$ are again trivial.
The solution of the field equations of $(r^2,\vartheta)$ is given as
\bsubeq \label{sol-r2r4-SD-522}
\begin{align}
\del_m r^2
\ &= \ 
- \frac{H}{K} \Big( \eps_{mn} \del^n \sigma_{2+}
- \frac{\Omega_2}{H} \del_m \wt{\vartheta}
\Big)
\, , \ls
K \ \equiv \ H^2 + (\Omega_2)^2
\, , \\
\del_m \vartheta 
\ &= \
\frac{H}{K} \Big( \eps_{mn} \del^n \wt{\vartheta}
+ \frac{\Omega_2}{H} \del_m \sigma_{2+}
\Big)
\, .
\end{align}
\esubeq
Substituting this into the Lagrangian, we eventually obtain the NLSM for the exotic $5^2_2$-brane:
\begin{align}
\Scr{L}_{5^2_2}
\ &= \ 
- \frac{H}{2} \Big\{ (\del_m \varrho)^2 + \varrho^2 (\del_m \vartheta_{\varrho})^2 \Big\}
- \frac{H}{2 K} \Big\{ (\del_m \sigma_{2+})^2 + (\del_m \wt{\vartheta})^2 \Big\}
\nn \\
\ & \ \ \ \ 
+ \frac{\Omega_2}{K} \eps^{mn} (\del_m \sigma_{2+}) (\del_n \wt{\vartheta}) 
+ \sqrt{2} \sum_a \eps^{mn} \del_m \big( (\vartheta - t_a^4) \mr{A}_{n,a} \big)
\, . \label{L-SD-522}
\end{align}
This is the model derived in \cite{Kimura:2013fda, Kimura:2015yla},
if we identify $\sigma_{2+}$ with $- y^2$.
Applying the smearing procedure and the reduction in section \ref{sect:IR-sdNLSM} to the GLSM (\ref{L-SD-comp-522}),
we again obtain the same result.
Indeed the GLSM (\ref{L-SD-comp-522}) is nothing but the starting model of \cite{Kimura:2013fda, Kimura:2015yla}.

We remark that the configuration of the target space of (\ref{L-SD-522}) is globally nongeometric.
This means that the function $\Omega_2$ is no longer single-valued with respect to the angular coordinate $\vartheta_{\varrho}$.
However, we stress that the monopole equation (\ref{monopole-eq}) is still valid. 
This is one of the features that the background is locally geometric.

\subsubsection*{``$5^3_2$-brane''}

Finally we try to investigate the NLSM exhibited in the fifth column in Table \ref{table:sdNLSM}.
Although the duality transformation of $\vartheta$ is straightforward, 
those of $(r^1,r^2)$ are difficult because they contribute to the functions $(H, \Omega_i)$.
Then we should again perform the smearing procedure along $(r^1,r^2)$.

Here we first smear the $r^2$ direction, and later the $r^1$ direction.
The setup of the smearing as follows.
First, we split the label $a$ into $m$ sectors as
\begin{align}
\{ a \} \ &= \ 
\big\{ \{ a_1 \} , \{ a_2 \} , \ldots , \{ a_j \} , \ldots , \{ a_m \} \big\}
\, , \ls
\sum_{a=1}^k 
\ = \ 
\sum_{j=1}^m \sum_{a_j=1}^{k_j}
\, , \ls
\sum_{j=1}^m k_j \ = \ k
\, .
\end{align}
Second, the FI parameters in the $j$-th sector are rewritten as
$(s_j^1, s_{a_j}^2, 0)$, where we set $t_a^3 = 0$ for simplicity.
Third, we compactify the $r^2$ direction of radius ${\cal R}_2$ and take the small limit ${\cal R}_2 \to 0$, as discussed in (\ref{smearing2-SD}).
This limit also implies $k_j \to \infty$ in each sector.
More precisely, we obtain the following forms in this limit:
\bsubeq \label{smearing2-SD-2}
\begin{align}
(R_{a_j})^2 \ &= \ 
(r^1 - s_j^1)^2 + (r^2 - s_{a_j}^2)^2 + (r^3)^2
\ = \ 
(\varrho_j)^2 + (r^3)^2
\, , \\
r^1 - s_j^1 \ &\equiv \ \varrho_j \, \cos \vartheta_j
\, , \ls
r^3 \ \equiv \ \varrho_j \, \sin \vartheta_j
\, , \\
H
\ &\equiv \ 
\frac{1}{g^2} + \sum_{j=1}^m \lim_{k_j\to\infty} \sum_{a_j=1}^{k_j} \frac{1}{\sqrt{2} \, R_{a_j}}
\ = \ 
\frac{1}{g^2} + \sum_{j=1}^m \sigma'' \log \frac{\Lambda_j}{\varrho_j}
\, , \ls
\sigma'' \ \equiv \ \frac{1}{\sqrt{2} \, \pi {\cal R}_2} 
\, , \\
\Omega_1
\ &\equiv \ 
\sum_{j=1}^m \lim_{k_j\to\infty} \sum_{a_j=1}^{k_j} \Omega_{1,a_j}
\ = \ 
0
\, , \\
\Omega_2
\ &\equiv \ 
\sum_{j=1}^m \lim_{k_j\to\infty} \sum_{a_j=1}^{k_j} \Omega_{2,a_j}
\ = \ 
\sum_{j=1}^m \sigma'' \vartheta_j
+ \text{(divergent part)}
\, . 
\end{align}
\esubeq
As mentioned before, this result still satisfies the Dirac monopole equation (\ref{monopole-eq}) non-trivially.
Fourth, we compactify the $r^1$ direction of radius ${\cal R}_1$, and
we set the FI parameters $s_j^1$ to
\begin{align}
s_j^1 \ &= \ 2 \pi {\cal R}_1 \, j \ \equiv \ s 
\, . 
\end{align}
Take the small limit ${\cal R}_1 \to 0$.
This limit indicates the large limit $m \to \infty$,
and the period $2 \pi {\cal R}_1$ is also infinitesimally small as $2 \pi {\cal R}_1 \equiv \d s$.
In this limit, the functions $(H, \Omega_i)$ are reduced to 
\bsubeq \label{smearing1-SD}
\begin{align}
H \ &= \ 
\frac{1}{g^2} + \lim_{m\to \infty} \sum_{j=1}^m \sigma'' \log \frac{\Lambda_j}{\varrho_j}
\ = \ 
\frac{1}{g^2} 
+ \lim_{L\to\infty} \frac{\sigma''}{2 \pi {\cal R}_1} \int_{-L}^L \d s
\log \frac{\Lambda_j}{\sqrt{s^2 + (r^3)^2}}
\nn \\
\ &= \ 
\frac{1}{g^2} + (\sigma' \sigma'') r^3
+ \text{(divergent part)}
\, , \ls
\sigma' \ \equiv \ \frac{1}{\sqrt{2} \, \pi {\cal R}_1}
\, , \\
\Omega_2 \ &= \ 
\lim_{m \to \infty} \sum_{j=1}^m \sigma'' \arctan \Big( \frac{r^3}{r^1 - s_j} \Big)
\ = \
\lim_{L\to\infty} \frac{\sigma''}{\sqrt{2} \, \pi {\cal R}_1} \int_{-L}^L \d s \arctan \Big( \frac{r^3}{r^1 - s} \Big)
\nn \\
\ &= \ 
0
\, . 
\end{align}
\esubeq
Then the function $K$ is reduced to $H^2$.
Moreover, all components of the KK-vector $\Omega_i$ vanish.
This reveals that the Dirac monopole equation (\ref{monopole-eq}) is no longer valid.

We continue to analyze the semi-doubled NLSM (\ref{L-SDN-comp}).
Since this model does not depend on non-derivative $(r^1, r^2)$ any more, 
we can perform the duality transformation via the integrating out the fields $(r^1,r^2,\phi'_{1-}, \vartheta)$:
\bsubeq \label{EOM-red-532-SD}
\begin{alignat}{2}
0 \ &= \ 
\del^m \Big\{
H \, \del_m r^1
- \eps_{mn} \, \del^n \sigma_{1-}
\Big\}
\, , &\ls
0 \ &= \ 
\del^m \Big\{
H \, \del_m r^2 
+ \eps_{mn} \, \del^n \sigma_{2+}
\Big\}
\, , \\
0 \ &= \ 
\del^m \Big\{
- \eps_{mn} r^3
\Big\}
\, , &\ls
0 \ &= \ 
\del^m \Big\{
H \, \del_m \vartheta
- \eps_{mn} \, \del^n \wt{\vartheta} 
\Big\}
\, .
\end{alignat}
\esubeq
Substitute them into the Lagrangian.
Then we obtain the final form:
\begin{align}
\Scr{L}
\ &= \ 
- \frac{H}{2} (\del_m r^3)^2 
- \frac{1}{2 H} \Big\{ (\del_m \sigma_{1-})^2 + (\del_m \sigma_{2+})^2 + (\del_m \wt{\vartheta})^2 \Big\}
\nn \\
\ & \ \ \ \ 
+ \sqrt{2} \sum_a \eps^{mn} \del_m \big( (\vartheta - t_a^4) \mr{A}_{n,a} \big)
\, . \label{L-SD-532?}
\end{align}
This NLSM tells us that there is no B-field on the target space geometry.
This is also caused by the disappearance of the KK-vector.
Thus the configuration is purely geometric.
However, the geometry is not Ricci-flat.
Then this configuration does not satisfy the field equations of ten-dimensional supergravity. 
Hence we conclude that the NLSM (\ref{L-SD-532?}) does not correctly capture the feature of the background of the exotic $5^3_2$-brane.

Go back to the GLSM (\ref{L-SD-comp-532}) which we obtained in the previous section.
This GLSM does not have isometry along the $(r^1,r^2)$ directions, neither.
Then we also apply the smearing procedures (\ref{smearing2-SD-2}) and (\ref{smearing1-SD}) to this GLSM after the low energy limit $e_a \to \infty$.
However, this again generates the trivial KK-vector.
Then the NLSM from the GLSM (\ref{L-SD-comp-532}) precisely coincides with (\ref{L-SD-532?}).

The lack of consistency with ten-dimensional supergravity comes from the breakdown of the Dirac monopole equation (\ref{monopole-eq}) 
via the smearing procedure.
This is because all of the physical information of the exotic $5^3_2$-brane is absorbed into the divergent part, though the duality transformation rule itself seems consistent.
Hence we confirm that the background is locally nongeometric, genuinely.

\section{Summary} 
\label{sect:summary}

In this paper we studied the duality transformation without isometry and applied it to the $\N=(4,4)$ GLSM for five-branes.

We first utilized complex (twisted) linear superfields which are transformed from (twisted) chiral superfields.
After constructing the dual Lagrangians, 
we replaced the complex (twisted) linear superfields with the sum of (twisted) chiral superfields.
Expanding these superfields in terms of the component fields,
we obtained the so-called ``semi-doubled'' Lagrangians which involve both the original and dual fields.
Compared with the duality transformation technique with isometry,
the procedure we demonstrated here has a strong benefit.
This is the dualization along any directions irrespective of the existence of isometry.
In particular, we can perform the duality transformation both the real and imaginary part of the original (twisted) chiral superfields.

Applying this technique to the analysis of the $\N=(4,4)$ GLSM and its dualized systems,
we obtained the ``semi-doubled'' GLSM for five-branes.
This model generates the conventional GLSMs for H-monopoles, KK-monopoles, and an exotic $5^2_2$-brane in quite a simple way.
In particular,
we also obtained the formal description of the semi-doubled GLSM for the exotic $5^3_2$-brane whose background is even locally nongeometric.
Taking the low energy limit of the semi-doubled GLSM,
we obtained the semi-doubled NLSM which also contains both the original and dual fields.
Integrating out a certain set of fields, we correctly derived the conventional NLSMs for the H-monopoles, the KK-monopoles and the exotic $5^2_2$-brane.

In the case of the model for the exotic $5^3_2$-brane, however, 
we found that the Dirac monopole equation, which governs the background structure of the five-branes, is broken down caused by the smearing procedure.
This is the feature of the nongeometric structure of the $5^3_2$-brane.
Hence we understood that the nongeometric structure can be traced if the Dirac monopole equation is non-trivially described even in the configuration of the exotic $5^3_2$-brane.
In order to realize this, we have to extend, at least, the semi-doubled NLSM to the doubled sigma model, double field theory, and/or $\beta$-supergravity which involve the kinetic terms of both the original and dual degrees of freedom.

\section*{Acknowledgements}

The author thanks
Yusuke Kimura,
Hisayoshi Muraki,
Takahiro Nishinaka
and 
Shin Sasaki
for valuable discussions and comments.
He also thanks the Yukawa Institute for Theoretical Physics at Kyoto University for hospitality during the YITP workshop on ``Developments in String Theory and Quantum Field Theory'' ({YITP-W-15-12}).
This work is supported by the MEXT-Supported Program for the Strategic Research Foundation at Private Universities ``Topological Science'' ({Grant No.~S1511006}). 
This is also supported in part by the Iwanami-Fujukai Foundation.

\begin{appendix}
\section*{Appendix}

\section{2D superfields}
\label{app:SF}

In this appendix we exhibit the conventions and the definition of various superfields in two-dimensional spacetime.
The notation is based on the previous work \cite{Kimura:2015yla}.

\subsection{Conventions}

Two-dimensional superspace is expanded by the conventional coordinates $x^m$ and the anti-commuting Grassmann coordinates $\theta^{\pm}$. 
They are complex Weyl spinors.
We define their hermitian conjugate as $\ol{\theta}{}^{\pm} = (\theta^{\pm})^{\dagger}$.
In order to classify superfields, we introduce the supercovariant derivatives $D_{\pm}$ and $\ol{D}{}_{\pm}$:
\begin{alignat}{2}
D_{\pm} \ &= \ 
\frac{\del}{\del \theta^{\pm}} 
- \I \ol{\theta}{}^{\pm} \del_{\pm}
\, , & \ls
\ol{D}{}_{\pm} \ &= \ 
- \frac{\del}{\del \ol{\theta}{}^{\pm}} 
+ \I \theta^{\pm} \del_{\pm}
\, . \label{sup-cov}
\end{alignat}
Here we used the light-cone coordinates $\del_{\pm} = \del_0 \pm \del_1$, 
where $\del_m = \frac{\del}{\del x^m}$.
It is also useful to define the integral measures of the Grassmann coordinates:
\bsubeq \label{f-measure-22}
\begin{gather}
\begin{alignat}{2}
\d^2 \theta \ &= \ 
- \half \, \d \theta^+ \, \d \theta^- 
\, , &\ls
\d^2 \ol{\theta} \ &= \ 
\half \, \d \ol{\theta}{}^+ \, \d \ol{\theta}{}^- 
\, , \\
\d^2 \wt{\theta} \ &= \ 
- \half \, \d \theta^+ \, \d \ol{\theta}{}^- 
\, , &\ls 
\d^2 \ol{\wt{\theta}} \ &= \ 
- \half \, \d \theta^- \, \d \ol{\theta}{}^+ 
\, , 
\end{alignat}
\\
\d^4 \theta \ = \ \d^2 \theta \, \d^2 \ol{\theta} 
\ = \ 
- \d^2 \wt{\theta} \, \d^2 \ol{\wt{\theta}} 
\ = \ - \frac{1}{4} \d \theta^+ \, \d \theta^- \, \d \ol{\theta}{}^+ \,
\d \ol{\theta}{}^- 
\, , \\
\int \! \d^2 \theta \, \theta \theta \ = \ 1 
\, , \ls
\int \! \d^2 \ol{\theta} \, \ol{\theta} \ol{\theta} \ = \ 1 
\, , \ls
\int \! \d^2 \wt{\theta} \, \theta^+ \ol{\theta}{}^- \ = \ \half 
\, , \ls
\int \! \d^2 \ol{\wt{\theta}} \, \theta^- \ol{\theta}{}^+ \ = \ \half
\, .
\end{gather}
\esubeq

\subsection{Various superfields}

Let us introduce various superfields in two dimensions.
First, we define a chiral superfield $X$ in such a way as
\begin{align}
0 \ &= \ \ol{D}{}_{\pm} X
\, . \label{def-ch}
\end{align}
This is an irreducible superfield. 
This means that the chiral superfield cannot decompose into any other superfields.
In the same way, we define another irreducible superfield, called a twisted chiral superfield $Y$:
\begin{align}
0 \ &= \ 
\ol{D}{}_+ D_- Y
\, . \label{def-Tch}
\end{align}
There is a real superfield $V = V^{\dagger}$ which carries a vector field.
In two dimensions, this is also described as a twisted chiral superfield $\Sigma$ in the following form:
\begin{align}
\Sigma \ &= \ 
\frac{1}{\sqrt{2}} \ol{D}{}_+ D_- V
\, . 
\end{align}
We note that this is an abelian vector superfield.

Relaxing the constraints in (\ref{def-ch}) and (\ref{def-Tch}), we can introduce reducible superfields.
We define a left semi-chiral superfield ${\mathbb X}$ and a right semi-chiral superfield ${\mathbb Y}$ as
\begin{align}
0 \ &= \
\ol{D}{}_+ {\mathbb X}
\, , \ls
0 \ = \ 
\ol{D}{}_- {\mathbb Y}
\, . \label{def-sc}
\end{align}
Furthermore, we define a complex linear superfield $L$ and a complex twisted linear superfield $\wt{L}$ as
\begin{align}
0 \ &= \ 
\ol{D}{}_+ \ol{D}{}_- L
\, , \ls
0 \ = \ 
\ol{D}{}_+ D_- \wt{L}
\, . \label{def-cls}
\end{align}
Due to the definition, a complex (twisted) linear superfield can be given as the sum of semi-chiral superfields:
\begin{align}
L \ &= \ {\mathbb X} + {\mathbb Y}
\, , \ls
\wt{L} \ = \ {\mathbb X} + \ol{\mathbb Y}
\, . \label{XY2L}
\end{align}

It is worth describing the expansion of superfields $(X, Y, V, L, \wt{L})$ by means of the Grassmann coordinates \cite{Kimura:2015yla, Kimura:2015cza}\footnote{We do not expand semi-chiral superfields which do not appear in the main part of this paper.}:
\bsubeq \label{expand}
\begin{align}
X \ &= \
\phi_X
+ \I \sqrt{2} \, \theta^+ \psi_{X+} 
+ \I \sqrt{2} \, \theta^- \psi_{X-}
+ 2 \I \, \theta^+ \theta^- F_X
\nn \\
\ & \ \ \ \ 
- \I \, \theta^+ \ol{\theta}{}^+ \del_+ \phi_X
- \I \, \theta^- \ol{\theta}{}^- \del_- \phi_X
+ \sqrt{2} \, \theta^+ \ol{\theta}{}^+ \theta^- \del_+ \psi_{X-}
+ \sqrt{2} \, \theta^- \ol{\theta}{}^- \theta^+ \del_- \psi_{X+}
\nn \\
\ & \ \ \ \
+ \theta^+ \theta^- \ol{\theta}{}^+ \ol{\theta}{}^- \del_+ \del_- \phi_X
\, , \label{chiralX} \\
Y \ &= \
\sigma_Y
+ \I \sqrt{2} \, \theta^+ \ol{\chi}{}_{Y+} 
- \I \sqrt{2} \, \ol{\theta}{}^- \chi_{Y-} 
+ 2 \I \, \theta^+ \ol{\theta}{}^- G_Y
\nn \\
\ & \ \ \ \ 
- \I \, \theta^+ \ol{\theta}{}^+ \del_+ \sigma_Y
+ \I \, \theta^- \ol{\theta}{}^- \del_- \sigma_Y
- \sqrt{2} \, \theta^- \ol{\theta}{}^- \theta^+ \del_- \ol{\chi}{}_{Y+}
- \sqrt{2} \, \theta^+ \ol{\theta}{}^+ \ol{\theta}{}^- \del_+ \chi_{Y-}
\nn \\
\ & \ \ \ \ 
- \theta^+ \theta^- \ol{\theta}{}^+ \ol{\theta}{}^- \del_+ \del_- \sigma_Y
\, , \label{tchiralY} \\
V \ &= \ 
- \theta^+ \ol{\theta}{}^+ (A_{0} + A_{1})
- \theta^- \ol{\theta}{}^- (A_{0} - A_{1})
- \sqrt{2} \, \theta^- \ol{\theta}{}^+ \sigma
- \sqrt{2} \, \theta^+ \ol{\theta}{}^- \ol{\sigma}
\nn \\
\ & \ \ \ \ 
- 2 \I \, \theta^+ \theta^-
\big( \ol{\theta}{}^+ \ol{\lambda}{}_{+} 
+ \ol{\theta}{}^- \ol{\lambda}{}_{-} \big)
+ 2 \I \, \ol{\theta}{}^+ \ol{\theta}{}^-
\big( \theta^+ \lambda_{+} + \theta^- \lambda_{-} \big)
+ 2 \, \theta^+ \theta^- \ol{\theta}{}^+ \ol{\theta}{}^- D_{V}
\, , \label{V} \\
L \ &= \ 
\phi_{L} 
+ \I \sqrt{2} \, \theta^+ \psi_{L+} 
+ \I \sqrt{2} \, \theta^- \psi_{L-} 
+ \I \sqrt{2} \, \ol{\theta}{}^+ \chi_{L+} 
+ \I \sqrt{2} \, \ol{\theta}{}^- \chi_{L-}
\nn \\
\ & \ \ \ \ 
+ \I \, \theta^+ \theta^- F_{L} 
+ \theta^+ \ol{\theta}{}^- G_{L} 
+ \theta^- \ol{\theta}{}^+ N_{L}
+ \theta^- \ol{\theta}{}^- A_{L=}
+ \theta^+ \ol{\theta}{}^+ B_{L\+}
\nn \\
\ & \ \ \ \ 
- \sqrt{2} \, \theta^+ \theta^- \ol{\theta}{}^+ \zeta_{L+}
- \sqrt{2} \, \theta^+ \theta^- \ol{\theta}{}^- \zeta_{L-}
+ \sqrt{2} \, \theta^+ \ol{\theta}{}^+ \ol{\theta}{}^- \del_+ \chi_{L-}
- \sqrt{2} \, \theta^- \ol{\theta}{}^+ \ol{\theta}{}^- \del_- \chi_{L+}
\nn \\
\ & \ \ \ \ 
+ \theta^+ \theta^- \ol{\theta}{}^+ \ol{\theta}{}^- \Big(
  \I \del_- B_{L\+} 
+ \I \del_+ A_{L=} 
- \del_+ \del_- \phi_{L} 
\Big)
\, , \label{CLS} \\
\wt{L} \ &= \ 
\wt{\phi}_{L} 
+ \I \sqrt{2} \, \theta^+ \wt{\psi}_{L+} 
+ \I \sqrt{2} \, \theta^- \wt{\psi}_{L-} 
+ \I \sqrt{2} \, \ol{\theta}{}^+ \wt{\chi}_{L+} 
+ \I \sqrt{2} \, \ol{\theta}{}^- \wt{\chi}_{L-}
\nn \\
\ & \ \ \ \ 
+ \I \, \theta^+ \theta^- \wt{F}_{L} 
+ \I \, \ol{\theta}{}^+ \ol{\theta}{}^- \wt{M}_{L}
+ \theta^+ \ol{\theta}{}^- \wt{G}_{L} 
+ \theta^- \ol{\theta}{}^- \wt{A}_{L=}
+ \theta^+ \ol{\theta}{}^+ \wt{B}_{L\+}
\nn \\
\ & \ \ \ \ 
- \sqrt{2} \, \theta^+ \theta^- \ol{\theta}{}^+ \del_+ \wt{\psi}{}_{L-}
- \sqrt{2} \, \theta^+ \theta^- \ol{\theta}{}^- \wt{\zeta}_{L-}
- \sqrt{2} \, \theta^+ \ol{\theta}{}^+ \ol{\theta}{}^- \wt{\lambda}_{L+}
+ \sqrt{2} \, \theta^- \ol{\theta}{}^+ \ol{\theta}{}^- \del_- \wt{\chi}{}_{L+}
\nn \\
\ & \ \ \ \ 
- \theta^+ \theta^- \ol{\theta}{}^+ \ol{\theta}{}^- 
\Big(
- \del_+ \del_- \wt{\phi}_{L} 
- \I \del_+ \wt{A}_{L=} + \I \del_- \wt{B}_{L\+} 
\Big) 
\, . \label{CtLS}
\end{align}
In the main part of this paper,
we describe a chiral superfield $\Phi$ in terms of its prepotential $C$ by
$\Phi = \ol{D}{}_+ \ol{D}{}_- C$.
This $C$ is unconstrained and complex. 
Its expansion is also exhibited as follows:
\begin{align}
C \ &= \ 
\phi_{c} 
+ \I \sqrt{2} \, \theta^+ \psi_{c+} 
+ \I \sqrt{2} \, \theta^- \psi_{c-} 
+ \I \sqrt{2} \, \ol{\theta}{}^+ \chi_{c+} 
+ \I \sqrt{2} \, \ol{\theta}{}^- \chi_{c-}
\nn \\
\ & \ \ \ \ 
+ \I \, \theta^+ \theta^- F_{c} 
+ \I \, \ol{\theta}{}^+ \ol{\theta}{}^- M_{c}
+ \theta^+ \ol{\theta}{}^- G_{c} 
+ \theta^- \ol{\theta}{}^+ N_{c}
%
+ \theta^- \ol{\theta}{}^- A_{c=}
+ \theta^+ \ol{\theta}{}^+ B_{c\+}
\nn \\
\ & \ \ \ \ 
- \sqrt{2} \, \theta^+ \theta^- \ol{\theta}{}^+ \zeta_{c+}
- \sqrt{2} \, \theta^+ \theta^- \ol{\theta}{}^- \zeta_{c-}
- \sqrt{2} \, \theta^+ \ol{\theta}{}^+ \ol{\theta}{}^- \lambda_{c+}
- \sqrt{2} \, \theta^- \ol{\theta}{}^+ \ol{\theta}{}^- \lambda_{c-}
\nn \\
\ & \ \ \ \ 
- 2 \theta^+ \theta^- \ol{\theta}{}^+ \ol{\theta}{}^- D_{c}
\, . \label{C}
\end{align}
\esubeq

\section{Duality transformations}
\label{app:DT}

In this appendix we briefly review the duality transformations with(out) isometry.

\subsection{General analysis}
\label{app:DT-gen}

We briefly argue the duality transformations with(out) isometry in two-dimensional $\N=(2,2)$ theory discussed by Grisaru, Massar, Sevrin and Troost \cite{Grisaru:1997ep}.
We begin with the most generic first order Lagrangian:
\begin{align}
\Scr{L} \ &= \
\int \d^4 \theta \, \Big\{ 
K (A, \ol{A}, B, \ol{B}) 
- A {\mathbb X} 
- \ol{A} \, \ol{\mathbb X}
- B {\mathbb Y}
- \ol{B} \, \ol{\mathbb Y}
\Big\}
\, , \label{gen-L}
\end{align}
where $A$ and $B$ are unconstrained complex prepotentials, 
and ${\mathbb X}$ and ${\mathbb Y}$ are semi-chiral superfields.
From now on, we impose various constraints on (\ref{gen-L}) and obtain the corresponding duality transformation rules.

First, if we impose $A = B$ on (\ref{gen-L}), then we obtain
\begin{align}
\Scr{L} \ &= \ 
\int \d^4 \theta \, \Big\{
K (A, \ol{A})
- A ({\mathbb X} + {\mathbb Y})
- \ol{A} (\ol{\mathbb X} + \ol{\mathbb Y})
\Big\}
\nn \\
\ &= \ 
\int \d^4 \theta \, \Big\{
K (A, \ol{A})
- A L - \ol{A} \, \ol{L}
\Big\}
\, , \label{gen-L-1}
\end{align}
where we used the relation (\ref{XY2L}).
If we integrate out $L$, we find the constraint $0 = \ol{D}{}_{\pm} A$.
This implies that $A$ becomes a chiral superfield.
Instead, if we integrate out $A$, we obtain the second order Lagrangian of $L$.
Thus it turns out that (\ref{gen-L-1}) is the first order Lagrangian which dualizes a chiral superfield to a complex linear superfield, and vice versa.
If we want to obtain the first order Lagrangian which dualizes a twisted chiral superfield to a complex twisted linear superfield and vice versa,
we just interchange the role of $B$ and $\ol{B}$:
\begin{align}
\Scr{L} \ &= \ 
\int \d^4 \theta \, \Big\{
K (A, \ol{A})
- A ({\mathbb X} + \ol{\mathbb Y})
- \ol{A} (\ol{\mathbb X} + {\mathbb Y})
\Big\}
\nn \\
\ &= \ 
\int \d^4 \theta \, \Big\{
K (A, \ol{A})
- A \wt{L} - \ol{A} \, \ol{\wt{L}}
\Big\}
\, . \label{gen-tL-1}
\end{align}

Second, assuming that the function $K$ in (\ref{gen-L-1}) depends only on $A + \ol{A}$, we rewrite (\ref{gen-L-1}) in the following way:
\begin{align}
\Scr{L}
\ &= \ 
\int \d^4 \theta \, \Big\{
K (A + \ol{A})
- \half (A + \ol{A}) (L + \ol{L})
- \half (A - \ol{A}) (L - \ol{L})
\Big\}
\, . \label{gen-L-2}
\end{align}
Integrating out $A - \ol{A}$, we obtain a new constraint $L = \ol{L}$.
This implies that $L$ is reduced to the sum of a twisted chiral and its conjugate $L = Y + \ol{Y}$. 
Then introducing a real prepotential $R = A + \ol{A}$, we obtain 
\begin{align}
\Scr{L}
\ &= \ 
\int \d^4 \theta \, \Big\{
K (R) - \half R (Y + \ol{Y})
\Big\}
\, . \label{gen-L-2'}
\end{align}
Integrating out $Y$, we find that $R$ becomes the sum of a chiral superfield and its conjugate.
Instead, integrating out $R$, we obtain the second order Lagrangian of $Y$.
Thus we understand that (\ref{gen-L-2'}) is the first order Lagrangian which dualizes a chiral superfield to a twisted chiral superfield \cite{Rocek:1991ps, Hori:2000kt}.

We can obtain a similar model when we assume that $K$ in (\ref{gen-tL-1}) depends only on $A + \ol{A}$:
\begin{align}
\Scr{L}
\ &= \ 
\int \d^4 \theta \, \Big\{
K (A + \ol{A})
- \half (A + \ol{A}) (\wt{L} + \ol{\wt{L}}) 
- \half (A - \ol{A}) (\wt{L} - \ol{\wt{L}}) 
\Big\}
\, . \label{gen-tL-2}
\end{align}
Integrating out $A - \ol{A}$, we obtain the constraint $\wt{L} = \ol{\wt{L}}$.
This indicates that $\wt{L}$ is the sum of a chiral superfield and its conjugate 
$\wt{L} = X + \ol{X}$.
Introducing a real prepotential $\wt{R} = A + \ol{A}$, we obtain
\begin{align}
\Scr{L}
\ &= \ 
\int \d^4 \theta \, \Big\{
K (\wt{R}) - \half \wt{R} (X + \ol{X})
\Big\}
\, . \label{gen-tL-2'}
\end{align}
This is the first order Lagrangian which dualizes a twisted chiral superfield to a chiral superfield and vice versa.
Because when we integrate out $X$, the prepotential $\wt{R}$ becomes the sum of a twisted chiral superfield and its conjugate.

\subsection{Twisted chiral with isometry}
\label{app:DT-tc}

We demonstrate the duality transformation of the Lagrangian (\ref{L-Theta}) established by Ro\v{c}ek and Verlinde \cite{Rocek:1991ps}, Hori and Vafa \cite{Hori:2000kt}, and Tong \cite{Tong:2002rq}.

\subsubsection*{Superfields}

First, we discuss in the superfield formalism.
Since (\ref{L-Theta}) has an isometry along the imaginary part of $\Theta$,
we introduce another first order Lagrangian different from (\ref{L-RtLV}):
\begin{align}
\Scr{L}_{B \Gamma V}
\ &= \ 
\int \d^4 \theta \, \Big\{
- \frac{1}{2 g^2} B^2
- 2 B V
- (\Gamma + \ol{\Gamma}) V
\Big\}
\, . \label{L-BGV}
\end{align}
Here $\Gamma$ is a chiral superfield, and $B$ is a real prepotential.
If we integrate out $\Gamma$, $B$ is constrained as $0 = \ol{D}{}_+ \ol{D}{}_- B = D_+ D_- B$.
This implies that $B$ is the sum of a twisted chiral superfield $Y$ and its conjugate, i.e., $B = Y + \ol{Y}$.
Substituting this into (\ref{L-BGV}) and identifying $Y$ with $\Theta$, we find the original Lagrangian (\ref{L-Theta}). 
Instead, if we evaluate the equation of motion for $B$ of the Lagrangian (\ref{L-BGV}), we obtain 
\begin{align}
0 \ &= \ 
- \frac{1}{g^2} B - (\Gamma + \ol{\Gamma}) - 2 V 
\, . 
\end{align}
Plugging this into (\ref{L-BGV}), we find the dual Lagrangian
\begin{align}
\Scr{L}_{B \Gamma V}
\ &= \
\frac{g^2}{2} \int \d^4 \theta \, \Big( \Gamma + \ol{\Gamma} + 2 V \Big)^2
\ \equiv \ 
\Scr{L}_{\Gamma V}
\, . \label{L-GV}
\end{align}
We also find the duality relation via the prepotential $B$:
\begin{align}
- \frac{1}{g^2} (\Theta + \ol{\Theta})
\ &= \ 
(\Gamma + \ol{\Gamma}) + 2 V
\, . \label{Theta2GammaV}
\end{align}

\subsubsection*{Component fields}

The descriptions (\ref{L-XYWV}) and (\ref{Theta2XYWV}) in the main part must contain the information of (\ref{L-GV}) and (\ref{Theta2GammaV}), respectively.
Then we express (\ref{L-Theta}), (\ref{L-GV}) and (\ref{Theta2GammaV}) in terms of the component fields.
This is because their explicit forms play an essential role in determining the reduction rule of redundant fields in (\ref{L-XYWV}).

Following (\ref{expand}), we expand the chiral superfield $\Gamma$ as follows:
\begin{align}
\Gamma \ &= \ 
\frac{1}{\sqrt{2}} (\gamma^3 + \I \gamma^4)
+ \I \sqrt{2} \, \theta^+ \zeta_+
+ \I \sqrt{2} \, \theta^- \zeta_-
+ 2 \I \, \theta^+ \theta^- G_{\Gamma}
+ \ldots 
\, ,
\end{align}
where ``$\ldots$'' involves derivative terms.
The expansion of the twisted chiral superfield $\Theta$ is exhibited in (\ref{Theta}).
The duality relation (\ref{Theta2GammaV}) provides a set of significant equations:
\begin{align}
r^3 \ &= \ - g^2 \gamma^3
\, , \ls
\pm \del_{\pm} \vartheta
\ = \ 
- g^2 D_{\pm} \gamma^4
\, , \ls
D_m \gamma^4 \ = \ \del_m \gamma^4 - \sqrt{2} A_m
\, . \label{Theta2Gamma-comp}
\end{align}
We notice that the relation between $\vartheta$ and $\gamma^4$ is described only with derivatives.
Furthermore, because of the twisting, the relative sign in front of the derivatives is different.
Substituting the above expansion into (\ref{L-Theta}) and (\ref{L-GV}), we obtain the explicit form of the two Lagrangians such as
\bsubeq \label{L-tc-app}
\begin{align}
\Scr{L}_{\Theta}
\ &= \ 
- \frac{1}{2 g^2} \Big\{
(\del_m r^3)^2
+ (\del_m \vartheta)^2
\Big\}
+ \sqrt{2} \Big\{
r^3 D_V + \vartheta F_{01}
\Big\}
+ \text{(other terms)}
\, , \label{L-Theta-comp-app} \\
\Scr{L}_{\Gamma V}
\ &= \ 
- \frac{1}{2 g^2} (\del_m r^3)^2
- \frac{g^2}{2} (D_m \gamma^4)^2
+ \sqrt{2} \, r^3 D_V
+ \text{(other terms)}
\, , \label{L-GV-comp-app}
\end{align}
\esubeq
where $F_{01} = \eps^{mn} \del_m A_n$ and $\eps^{01} = - \eps^{10} = + 1$,
and we simply ignore the detail of the contribution from auxiliary fields and fermions.

\end{appendix}

}
\end{document}